\documentclass[aoas]{imsart}

\usepackage[dvipsnames]{xcolor}
\definecolor{DarkBlue}{rgb}{0,0,.545}
\definecolor{DarkRed}{rgb}{0.5,0,0}
\RequirePackage{amsthm,amsmath,amsfonts,amssymb}
\RequirePackage[authoryear]{natbib}
\RequirePackage[colorlinks,citecolor=DarkBlue, urlcolor=DarkRed]{hyperref}
\RequirePackage{graphicx}

\usepackage{subcaption}
\usepackage[font=footnotesize, labelfont=sc, textfont={footnotesize, it}]{caption}
\usepackage{bbm}

\usepackage{enumitem}

\startlocaldefs
\theoremstyle{plain}

\newtheorem{theorem}{Theorem}[section]

\theoremstyle{definition}
\newtheorem{definition}[theorem]{Definition}

\newtheorem{proposition}[theorem]{Proposition}

\endlocaldefs

\begin{document}

\begin{frontmatter}
\title{Bayesian nonparametric modeling of dynamic pollution clusters through an autoregressive logistic-beta Stirling-gamma process}
\runtitle{A Bayesian nonparametric AR-LB-SG process for dynamic clustering}

\begin{aug}
\author[A]{\fnms{Santiago}~\snm{Marin}\orcid{0009-0006-0086-3721}},
\author[A]{\fnms{Bronwyn}~\snm{Loong}\orcid{0000-0002-1409-8892}}
\and
\author[A,B]{\fnms{Anton H.}~\snm{Westveld}\orcid{0000-0003-1671-0931}}
\address[A]{Research School of Finance, Actuarial Studies and Statistics, Australian National University\printead[presep={.\\ }]{}}

\address[B]{Department of Mathematics and Statistics,
Virginia Commonwealth University\printead[presep={.\ }]{}}
\end{aug}

\begin{abstract}
Fine suspended particulates (FSP), commonly known as PM\textsubscript{2.5}, are among the most harmful air pollutants, posing serious risks to population health and environmental integrity. As such, accurately identifying clusters of air-quality monitoring stations exhibiting similar FSP levels is essential for guiding targeted interventions and localized public-health response measures. This task, however, is notably nontrivial as FSP levels may depend on various regional and temporal factors, which should be incorporated in the modeling process. Thus, we capitalize on Bayesian nonparametric dynamic clustering ideas, in which clustering structures may be influenced by complex dependencies. Existing implementations of dynamic clustering, however, rely on copula-based dependent Dirichlet processes (DPs), presenting considerable computational challenges for real-world deployment. With this in mind, we propose a more efficient alternative for dynamic clustering by incorporating the novel ideas of logistic-beta dependent DPs. We also adopt a Stirling-gamma prior---a novel distribution family---on the concentration parameter of our underlying DP, easing the process of incorporating prior knowledge into the model. Efficient computational strategies for posterior inference are also presented. We apply our proposed method to identify dynamic clusters of air-quality monitoring stations across Chile and demonstrate its superior performance over existing approaches.
\end{abstract}


\end{frontmatter}

\section{Introduction}\label{sec:intro}
Fine suspended particulates (FSP), commonly known as $\textnormal{PM}_{2.5}$, are the smaller respirable suspended particulates, and refer to those airborne microparticulates with a nominal aerodynamic diameter smaller or equal than 2.5 micrometres ($\mu\textnormal{m}$). Due to their notably fine dimensions, these particulates can penetrate and settle in the deepest parts of the lungs, the blood stream \citep{fsp_blood}, the brain \citep{fsp_brain}, and other body organs, increasing the risk of lung morbidity \citep{DONALDSON1998}, respiratory and cardiovascular diseases \citep{schwartz2000particles, Beijing_hospitals}, as well as different types of cancer \citep{hill2023lung, White_2023_breastcancer}. In fact, even a short-term exposure to large concentrations of FSP can significantly deteriorate people's health \citep{Deb_Tsay_Sinica_2019}. What is more, FSP may also lead to visibility impairments in the air and influence the severity and frequency of extreme climate events \citep{ZHANG201511, liang2015assessing}. It is not a surprise, then, that public and environmental authorities---from various countries around the world---are now investing in technologies and infrastructure to monitor concentration levels of various air pollutants, including FSP.
\par
With this in mind, thorough this article, we aim to accurately identify time-varying clusters of air-quality monitoring stations that exhibit similar FSP levels. Such clusters can aid public health and environmental authorities determine where and when considerable pollution episodes occur, which could then guide targeted interventions, localized public-health responses, environmental surveillance, and quality-assurance strategies. For example, identifying clusters of stations with similar FSP patterns may reveal: (a) common emission sources and (b) groups of individuals exposed to similar FSP concentration levels---and therefore, similar health risks. It is also well-known that model-based clustering algorithms can be used for outlier detection (see, e.g., \cite{newcomb1886generalized} for an early reference). Thus, one could also detect anomalous stations (e.g., a station that abruptly changes clusters across time), which may indicate instrument malfunction, data-quality issues, or extreme pollution events. Such information can then be used to develop more targeted monitoring, management, and mitigation strategies.
\par
\begin{figure*}[!htp]
\centering
\hspace{-0.75cm}
\begin{subfigure}{0.32\linewidth}
    \includegraphics[width=\linewidth]{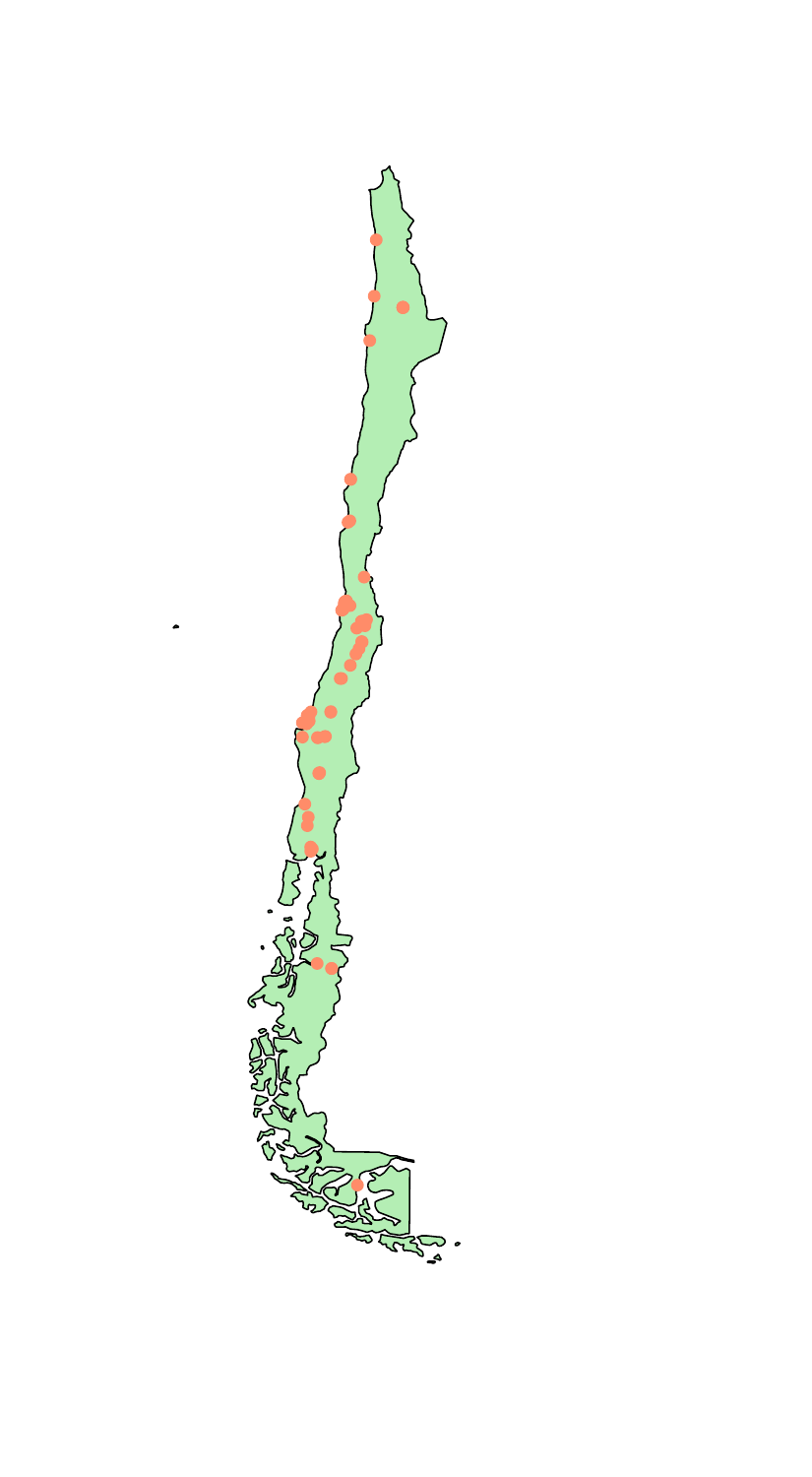}
    \caption{Monitoring stations}
\end{subfigure}
\hspace{-0.80cm}
\begin{subfigure}{0.75\linewidth}
    \includegraphics[width=\linewidth]{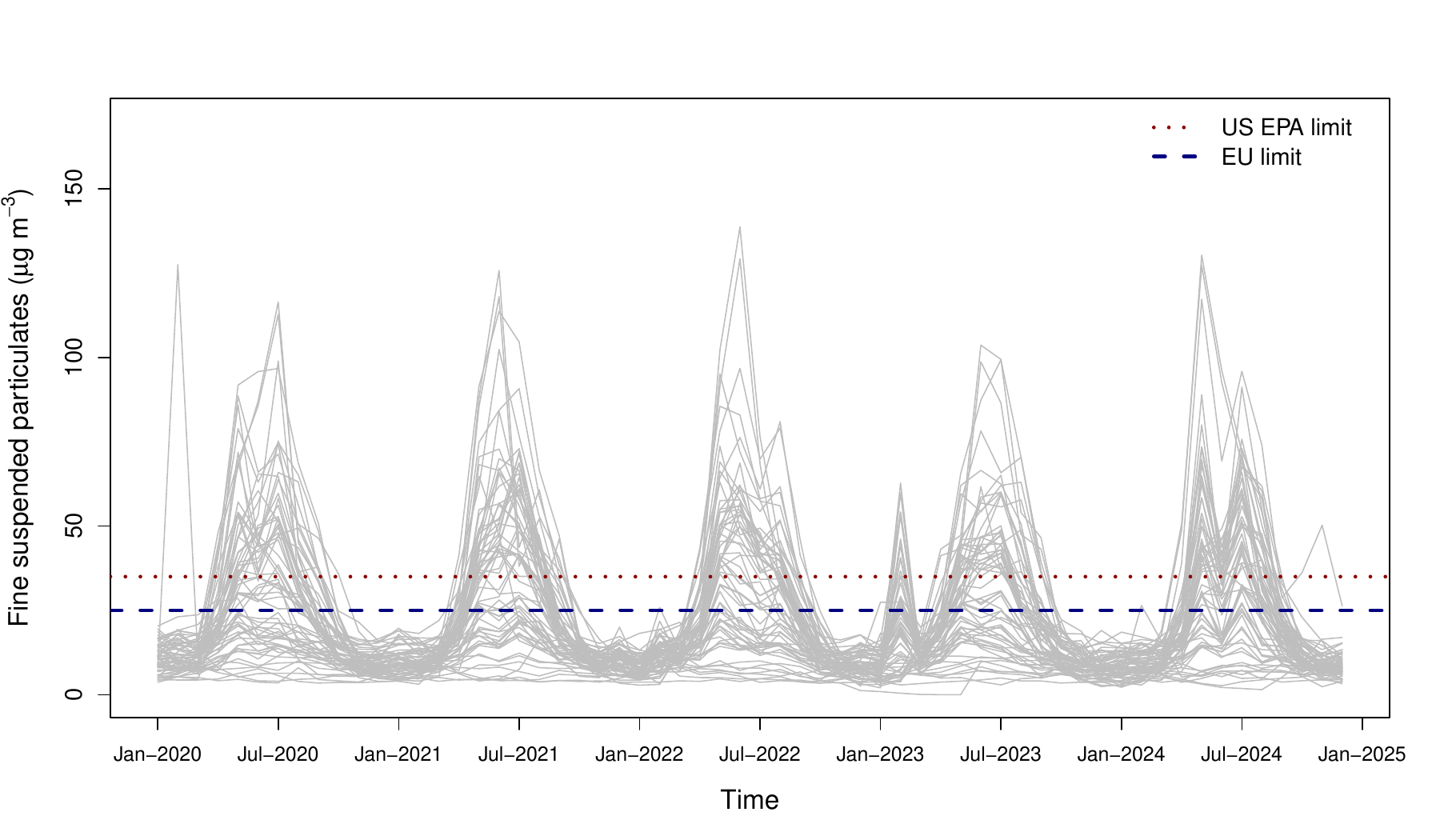}
    \vspace{0.8cm}
    \caption{Monthly average of FSP}
\end{subfigure}
\caption{\textsc{(a)} Locations of the considered 64 monitoring stations across continental Chile. \textsc{(b)} Evolution of the monthly averaged FSP concentration levels from each monitoring station. Gray lines correspond to monitoring stations. The red dotted lines and blue dashed lines denote the healthy FSP concentration limits established by the U.S. Environmental Protection Agency (EPA) and the European Union, respectively.}
\label{fig:stations}
\end{figure*}

In this particular study, we employ data from 64 air-quality monitoring stations across continental Chile (i.e., excluding its insular territories), obtained from the Chilean Ministry of the Environment's \textit{National Air Quality Information System} (SINCA, by its Spanish acronym). The dataset covers from January 2020 to December 2024 (i.e., 60 months), with stations distributed nationwide, although most of them are located in the central region. Figure \ref{fig:stations} presents the geographical locations of such monitoring stations---panel (\textsc{a}), as well as the evolution of the monthly-averaged FSP concentration levels---panel (\textsc{b}). Figure \ref{fig:stations} also displays the healthy FSP concentration limits established by the U.S. Environmental Protection Agency (EPA), set at 35 $\mu\text{g}\;\text{m}^{-3}$, and by the European Union, set at 25 $\mu\text{g}\;\text{m}^{-3}$ \citep{liang2015assessing}.

\begin{figure*}
\begin{center}
\includegraphics[width=0.8\linewidth]{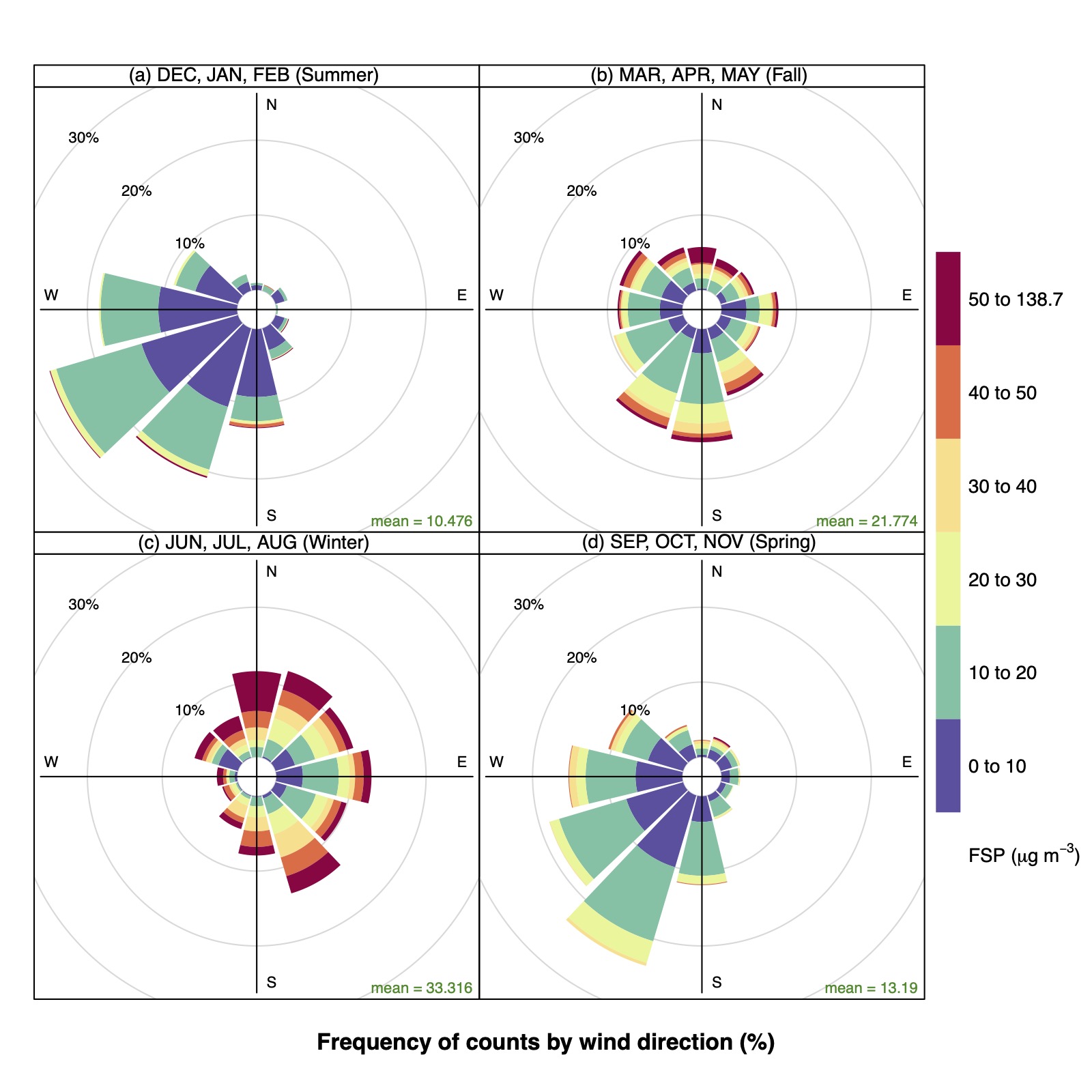}
\end{center}
\caption{Pollution roses depicting seasonal FSP concentration levels and wind-directional patterns across Chilean monitoring stations.}
\label{fig:seasonal_wind}
\end{figure*}

It is clear, from Figure \ref{fig:stations}, that a substantial proportion of monitoring stations across Chile have consistently recorded FSP concentration levels far above the recommended healthy limits over prolonged periods of time, posing a serious risk to the Chilean population. This illustrates the need to accurately identify time-varying clusters of air-quality monitoring stations across Chile, which can then be translated into targeted actions aimed at protecting the Chilean population. Achieving this, however, is notably nontrivial as FSP concentration levels depend on regional and temporal factors---such as topographical, seasonal climatic, and land cover variables (see e.g., \cite{haas1995local}, \cite{Kibria_jasa_2002}, \cite{Sahu2005Krig}, \cite{liang2015assessing} or \cite{Deb_Tsay_Sinica_2019}, just to name a few). For instance, Figure \ref{fig:seasonal_wind} illustrates how FSP concentration levels in Chile vary with seasonal and wind-directional patterns, where it is clear that periods of wintertime north-easterly winds coincide with greater FSP levels. Therefore, any appropriate model aiming to identify latent clusters based on FSP levels should account for these intricate dynamics. 

\begin{figure*}
\begin{center}
\includegraphics[width=\linewidth]{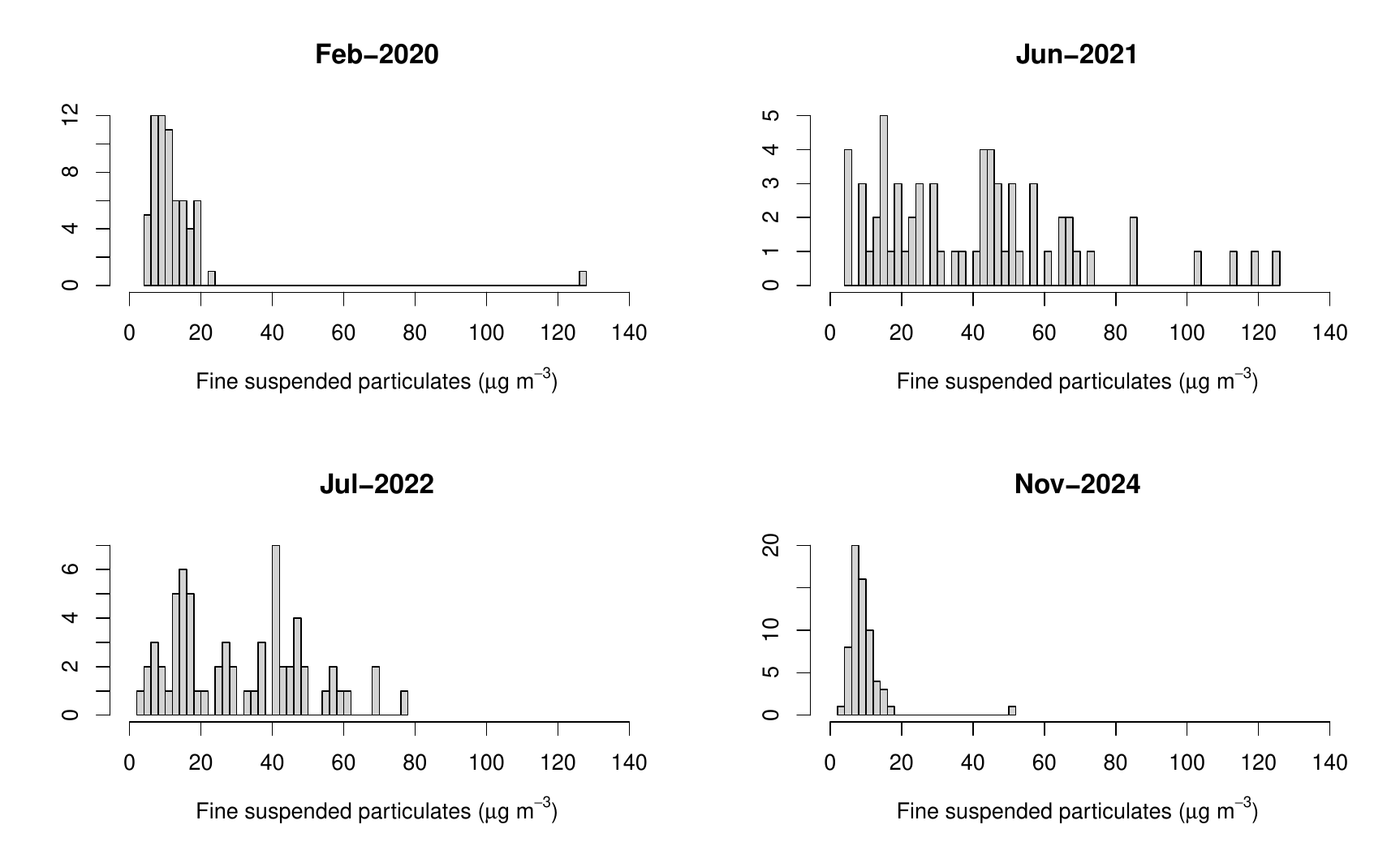}
\end{center}
\caption{Histograms of FSP concentration levels recorded at Chilean monitoring stations across four different time points.}
\label{fig:chile_hists}
\end{figure*}

Considering this, we capitalize on the flexibility of Bayesian nonparametric methods. In particular, we build on the ideas of \textit{dynamic clustering} from \cite{DeIorio2023_AR_DPMs} (also known as \textit{multi-view clustering} \citep{Franzolini_multiView}), in which the number of clusters and the cluster allocations are allowed to change over time, while accounting for complex temporal and regional dependencies in the data. To further highlight the need for dynamic clustering, Figure \ref{fig:chile_hists} presents histograms of FSP concentration levels at four different time points using the Chilean dataset. It is clear, from Figure \ref{fig:chile_hists}, that clusters do split and merge over time. More precisely, we can observe that in February 2020 and November 2024, most of the FSP readings are near one another, with very few readings far in the upper tail. During June 2021 and July 2022, on the other hand, FSP readings exhibit a much wider dispersion, illustrating dynamic shifts in the data generating mechanism.

Existing implementations of dynamic clustering, however, rely on copula-based dependent Dirichlet processes, which, despite their well-known versatility, exhibit considerable computational challenges for real-world deployment. With this in mind, we incorporate the novel ideas of logistic-beta dependent Dirichlet processes from \cite{logistic_beta_2025} as they provide a more efficient alternative to copula-based dependent Dirichlet processes, while preserving all the necessary capabilities to perform dynamic clustering.
\par
Moreover, it is widely acknowledged that eliciting an informative Dirichlet process prior is far from simple \citep{escobar1994estimating, lijoi2007controlling}. To exacerbate these challenges, commonly used prior distributions on the concentration parameter of a Dirichlet process---such as the gamma distribution---lead to analytically intractable results, preventing a clear and intuitive incorporation of prior beliefs into the model specification. Thus, we also incorporate a Stirling-gamma prior \citep{zito2023bayesian}, a novel distribution family, on the concentration parameter of our underlying Dirichlet process. Ultimately, this facilitates the elicitation of an informative prior and makes the workflow more transparent and straightforward. As such, we call our proposed method the autoregressive logistic-beta Stirling-gamma process.
\par
The use of Bayesian nonparametric models to analyze environmental quality data is not new. For instance, \cite{Sahu2005Krig} employed a Bayesian kriged Kalman filter to forecast FSP levels across New York City; \cite{atmos11111233} employed Bayesian additive regression trees to predict individual components of FSP across California; \cite{arbel2016bayesian} used a copula-based dependent Dirichlet process to assess the effect of a fuel spill in Antarctica on species diversity; \cite{gutierrez2016time} employed a time-dependent Dirichlet process mixture to estimate a time-varying density and the probability of air pollutants exceeding an arbitrary threshold, at a given time point, using FSP data from the metropolitan region of Santiago, Chile. This differs from our work as the authors focus on dynamic density estimation rather than dynamic clustering. Moreover, as discussed in \cite{DeIorio2023_AR_DPMs}, the method introduced in this article accommodates both dynamic density estimation and dynamic clustering, highlighting its advantages. Closer to our method, on the other hand, is the study of \cite{Argiento_pollution_2026}, which uses spatial product partition models to cluster air-quality monitoring stations exhibiting similar $\textnormal{PM}_{10}$ dynamics across Northern Italy, and the work of \cite{page2022dependent}, which introduces a dependent random partition model for dynamic clustering and apply it to identify of clusters of air-quality monitoring stations in rural Germany. As demonstrated later, however, our proposed method achieves a comparable and, in some settings, superior performance.
\par
The remainder of this article is then organized as follows. Section \ref{sec:background} reviews autoregressive Dirichlet process mixtures, the logistic-beta process, and the Stirling-gamma distribution. In Section \ref{sec:AR_SGP}, we formally introduce our model, illustrate its analytical properties, and present details of the computational strategies employed for posterior inference. Simulation exercises are carried out in Section \ref{sec:simulations}. In section \ref{sec:chile_data}, we apply our model to identify dynamic clusters of air-quality monitoring stations exhibiting similar FSP levels across continental Chile. We conclude with a discussion in Section \ref{sec:discussion}.

\section{Background}
\label{sec:background}
\subsection{Copula-based autoregressive Dirichlet processes}
\label{subsec:DPMs}
Discrete random probability measures---and the latent partitions induced by them---have been at the center of Bayesian nonparametric clustering. Due to its simplicity and analytical tractability, Ferguson's Dirichlet process \citep{Ferguson_DP} has been arguably the most popular and widely used prior law in Bayesian nonparametric analysis. More formally, we denote $G\sim\textnormal{DP}(\alpha, G_0)$ whenever a random probability measure, $G$, follows a  Dirichlet process (DP) with concentration parameter $\alpha>0$ and  centering measure $G_0$ defined over the metric space $\Theta$. An interesting property of the DP, nonetheless, is that it generates random probability measures that are almost surely discrete (see also \cite{Blackwell_discrete} and \cite{blackwell1973ferguson}). Based on this almost sure discreteness, \cite{sethuraman1994constructive} consequently showed that $G$ can then be expressed as $G(\cdot)=\sum_{k\in\mathbb{N}}w_{k}\delta_{\zeta_{k}}(\cdot)$, where $\delta_{A}$ denotes the Dirac measure at $A$, $\{\zeta_{k}\}_{k\in\mathbb{N}}\overset{\text{iid}}{\sim}G_{0}$, $w_k = \nu_{k}\prod_{l<k}(1 - \nu_{l})$, $\{\nu_{k}\}_{k\in\mathbb{N}} \overset{\text{iid}}{\sim} \text{Beta}(1,\alpha)$, $\{\zeta_{k}\}_{k\in\mathbb{N}}\perp\{\nu_{k}\}_{k\in\mathbb{N}}$, and $\mathbb{N}=\{1,2,3,\dots\}$. This is known as the \textit{stick-breaking} construction of the DP.
\par
Building upon this stick-breaking construction, \cite{DeIorio2023_AR_DPMs} formulated a time-dependent sequence of random probability measures, $\{G_{t}\}_{t\in\mathbb{N}}$, where $\{G_{t}\}_{t\in\mathbb{N}}$ has an autoregressive (AR) structure and, for any $t\in\mathbb{N}$, $G_t\sim\text{DP}(\alpha, G_0)$. To do so, the authors employ a latent AR process of order one, denoted by $\boldsymbol{\epsilon}=\{\epsilon_{t}\}_{t\in\mathbb{N}}\sim\text{AR}(1;\psi)$, where $\epsilon_1\sim N(0,1)$ and, for $t\in\{2,3,\dots\}$, $\epsilon_{t}=\psi\epsilon_{t-1} + \eta_{t}$, with $|\psi|<1$ and $\{\eta_t\}_{t\in\mathbb{N}}\overset{\text{iid}}{\sim}N(0, 1 - \psi^{2})$ so that $\epsilon_{t}\sim N(0,1)$. Additionally, the authors also make use of a copula-based transformation of a Gaussian AR process (see e.g., \cite{Guolo_google}). More precisely, let $H(\cdot; a,b)$ be the cumulative distribution function (CDF) of a beta-distributed random variable with parameters $a$ and $b$. If $\epsilon\sim N(0,1)$, then $X=H^{-1}(\Phi(\epsilon);a,b)$ would be a beta-distributed random variable with parameters $a$ and $b$, where $\Phi$ denotes the standard normal CDF. Hence, the authors consider the sequence $\{\boldsymbol{\epsilon}_k\}_{k\in\mathbb{N}}\overset{\text{iid}}{\sim}\text{AR}(1;\psi)$ and let, for $t,k\in\mathbb{N}$,
\begin{equation}
    \label{eq:prob_int_transform}
    \begin{split}
        \nu_{tk} & = H^{-1}(\Phi(\epsilon_{tk}); 1, \alpha)  = 1 - (1 - \Phi(\epsilon_{tk}))^{1/\alpha},
    \end{split}
\end{equation}
such that, for any $t$, $\{\nu_{tk}\}_{k\in\mathbb{N}}\overset{\text{iid}}{\sim}\text{Beta}(1,\alpha)$. Moreover, due to the autoregressive structure of $\boldsymbol{\epsilon}_k$, $\nu_{tk}$ would depend on $\nu_{(t-1)k}$, with $\psi$ controlling the dependence among the $\nu_{tk}$'s.
\par
Following the stick-breaking construction of the DP, one can then let, for $t\in\mathbb{N}$,
\begin{equation}
    \label{eq:DeIorio_DDP}
    G_{t}(\cdot) = \sum_{k\in\mathbb{N}}w_{tk}\delta_{\zeta_{k}}(\cdot),
\end{equation}
where $\{\zeta_{k}\}_{k\in\mathbb{N}}\overset{\text{iid}}{\sim}G_{0}$, $w_{tk}= \nu_{tk}\prod_{l<k}(1 - \nu_{tl})$, and $\nu_{tk}$ is constructed as in \eqref{eq:prob_int_transform} independent of $\{\zeta_{k}\}_{k\in\mathbb{N}}$. Since $\{\nu_{tk}\}_{k\in\mathbb{N}}\overset{\text{iid}}{\sim}\text{Beta}(1,\alpha)$, then, for any $t\in\mathbb{N}$, $G_t\sim\text{DP}(\alpha, G_0)$. Moreover, since the $\nu_{tk}$'s have an AR structure, the sequence $\{G_t\}_{t\in\mathbb{N}}$ also has an AR structure \citep{DeIorio2023_AR_DPMs}. This is known as a \textit{copula-based autoregressive Dirichlet process}, denoted by $\{G_{t}\}_{t\in\mathbb{N}}\sim\text{AR-DP}(\psi, \alpha, G_0)$, which is a special case of a dependent DP \citep{maceachern2000dependent, barrientos_etal2012, quintana2022dependent} and a telescopic clustering model \citep{Franzolini_multiView}.
\par
That being said, copula-based dependent DPs have been widely used in the Bayesian nonparametric literature; some other notable examples include the studies by \cite{rodriguez2010latent}, \cite{rodriguez2011nonparametric}, and \cite{arbel2016bayesian}, just to name a few. Other examples of time-dependent discrete random measures can be found in \cite{caron2007}, \cite{caron2008}, \cite{dunson2008kernel}, \cite{taddy2010autoregressive}, \cite{griffin2011stick}, \cite{xiao2015modeling}, \cite{caron2017}, and \cite{deyoreo2018modeling}. More recently, \cite{Grazian_temporal_2025} proposed a DP in which the stick-breaking weights incorporate spatio-temporal dependencies in order to model and predict dynamic surfaces, rather than performing dynamic clustering.

\subsection{Logistic-beta dependent Dirichlet processes}
\label{sec:LB_DP}
Copula-based formulations of dependent DPs, while flexible, often yield notable computational challenges. For instance, the full conditional distribution of the stick-breaking weights in the AR-DP from \cite{DeIorio2023_AR_DPMs} is given by a Bayesian nonlinear temporal dynamic model, so the authors resort to a computationally expensive particle Markov chain Monte Carlo algorithm \citep{Andrieu_PMCMC} to generate random draws from such a distribution. \cite{logistic_beta_2025} on the other hand, recently introduced the \textit{logistic-beta process}, a novel stochastic process which enables efficient posterior computations in dependent DPs through a normal variance-mean mixture representation of the stick-breaking weights. What follows is a review of the applications of the logistic-beta process in Bayesian nonparametric inference.
\par 
As in \cite{logistic_beta_2025}, a random variable $\epsilon\in\mathbb{R}$ is said to follow a univariate logistic-beta distribution with shape parameters $a_\epsilon,b_\epsilon>0$ if its density function is given by
\begin{equation}
    \label{eq:univariate_logistic_beta}
    p(\epsilon)=\frac{1}{\textnormal{B}(a_\epsilon,b_\epsilon)}\left(\frac{1}{1+\exp(-\epsilon)}\right)^{a_{\epsilon}}\left(\frac{\exp(-\epsilon)}{1+\exp(-\epsilon)}\right)^{b_{\epsilon}}, 
\end{equation}
where $\textnormal{B}:\mathbb{R}_+\times\mathbb{R}_+\rightarrow\mathbb{R}_+$ denotes the usual beta function. Thus, setting $a_\epsilon=b_\epsilon=1$ results in a standard logistic distribution. Notably, though, applying the logistic transformation [i.e., $x\mapsto\textnormal{logit}^{-1}(x)=1/(1+\exp(-x))$] to a logistic-beta random variable, $\epsilon$, yields a beta-distributed random variable with parameters $a_\epsilon$ and $b_\epsilon$. In other words, $\textnormal{logit}^{-1}(\epsilon)\sim\textnormal{Beta}(a_\epsilon,b_\epsilon)$.
\par
Conveniently, we can express the logistic-beta density function in \eqref{eq:univariate_logistic_beta} as a normal variance-mean mixture with a P\'olya mixing density \citep{Barndorff_mixtures_1982}. More formally, the density function in \eqref{eq:univariate_logistic_beta} can be written as
\begin{equation*}
    p(\epsilon)=\int_{\mathbb{R}_+}\phi(\epsilon;\,0.5\lambda(a_\epsilon-b_\epsilon),\lambda)\pi_{\textnormal{Polya}}(\lambda;\,a_\epsilon,b_\epsilon)\textnormal{d}\lambda,
\end{equation*}
where $\phi(\cdot;\mu,\sigma^2)$ denotes the density function of a normally-distributed random variable with mean $\mu$ and variance $\sigma^2$, while $\pi_{\textnormal{Polya}}(\cdot;a,b)$ denotes the density function of a P\'olya-distributed random variable with shape parameters $a$ and $b$. 
\par
Based on the above mixture representation, \cite{logistic_beta_2025} formulated a multivariate logistic-beta distribution that has the same univariate logistic-beta marginals as in \eqref{eq:univariate_logistic_beta}.
\begin{definition}[The $d$-dimensional multivariate logistic-beta distribution \citep{logistic_beta_2025}] 
Let $\boldsymbol{\Psi}$ be a $d\times d$ positive semidefinite correlation matrix. Then, a random vector $\boldsymbol{\epsilon}\in\mathbb{R}^d$ is said to follow a $d$-dimensional multivariate logistic-beta  distribution with shape parameters $a_\epsilon,b_\epsilon>0$ and correlation parameter $\boldsymbol{\Psi}$, denoted as $\boldsymbol{\epsilon}\sim\textnormal{LB}(a_\epsilon,b_\epsilon,\boldsymbol{\Psi})$, if
    \begin{equation*}
        \boldsymbol{\epsilon}|\lambda\sim N_{d}(0.5\lambda(a_\epsilon-b_\epsilon)\mathbf{1}_d,\lambda\boldsymbol{\Psi}),\qquad\lambda\sim\textnormal{Polya}(a_\epsilon,b_\epsilon),
    \end{equation*}
where $\textnormal{Polya}(a,b)$ denotes the P\'olya distribution with shape parameters $a$ and $b$, and $\mathbf{1}_{d}=(1,\dots,1)'\in\mathbb{R}^{d}$ denotes the $d$-dimensional column vector of ones.
\end{definition}
\par
The above  multivariate logistic-beta distribution enables us to effectively model dependent DPs through a logistic-beta process. More formally, let $\mathcal{R}:\mathcal{X}\times\mathcal{X}\rightarrow[-1,1]$ be a positive semidefinite correlation kernel with $\mathcal{R}(x, x) = 1$, for all $x\in\mathcal{X}$. Then, we say that $\left\{\epsilon(x)\in\mathbb{R}:x\in\mathcal{X}\right\}$ follows a logistic-beta process with shape parameters $a_\epsilon,b_\epsilon>0$ and correlation kernel $\mathcal{R}$, denoted as $\textnormal{LBP}(a_\epsilon,b_\epsilon,\mathcal{R})$, if every finite collection $\left\{\epsilon(1),\dots,\epsilon(d)\right\}'$ follows a $d$-dimensional multivariate logistic-beta distribution with shape parameters $a_\epsilon,b_\epsilon$, and correlation parameter $\mathbf{R}$ so that $(\boldsymbol{\mathbf{R}})_{i,i'}=\mathcal{R}(i,i')$ \citep{logistic_beta_2025}. Therefore, if $\left\{\epsilon_k(x)\in\mathbb{R}:x\in\mathcal{X}\right\}\sim\textnormal{LBP}(1, \alpha,\mathcal{R})$, then, for all $x\in\mathcal{X}$, $\textnormal{logit}^{-1}(\epsilon_k(x))\sim\textnormal{Beta}(1, \alpha)$ so that
\begin{equation*}
    \sum_{k\in\mathbb{N}}\left(\textnormal{logit}^{-1}(\epsilon_k(x))\prod_{l<k}\left[1-\textnormal{logit}^{-1}(\epsilon_l(x))\right]\right) \overset{\text{a.s.}}{=} 1,
\end{equation*}
defining proper stick-breaking weights \citep{sethuraman1994constructive, Ishwaran_James_2001_stick}, which depend on $x\in\mathcal{X}$ via $\epsilon_k(x)$. 
\par
It is clear, then, that the main advantage of logistic-beta dependent DPs is that one can sample from the full conditional distribution of the stick-breaking weights using the existing machinery to generate multivariate normally distributed random draws (see e.g., \cite{Rue_2001}). Despite these computational benefits, though, most of the literature on logistic-beta dependent DPs has been largely limited to nonparametric density regression tasks. Thus, we incorporate this novel idea into the design of a dynamic clustering algorithm.

\subsection{Bayesian nonparametric prior elicitation and the Stirling-gamma distribution}
\label{Sec:SGprior}

It is widely acknowledged that eliciting an informative DP prior is far from simple. For instance, clustering results induced by DPMs are notably sensitive to the choice of the concentration parameter $\alpha$ \citep{escobar1994estimating, lijoi2007controlling}. On top of this, \cite{miller2013simple, miller2014inconsistency} showed that the posterior distribution of the number of clusters induced by a DPM with a fixed $\alpha$ may not asymptotically concentrate around the true number of clusters. \cite{ascolani2023clustering}, however, showed that treating $\alpha$ as random---through a prior distribution, $p(\alpha)$---may yield consistent clustering results. So far, the most popular prior on $\alpha$ has been the gamma distribution \citep{escobar1995bayesian}, which was employed in the autoregressive DPM from \cite{DeIorio2023_AR_DPMs}. That being said, the lack of closed-form analytical results has prevented a transparent and straightforward incorporation of prior beliefs  with a gamma prior \citep{zito2023bayesian}.
\par
More formally, following \cite{Antoniak_DPM}, if $G|\alpha\sim\textnormal{DP}(\alpha,G_0)$ with a fixed $\alpha$, the probability of the random partition $\Pi_n$ induced by the a.s. discreteness of $G$ would be given by
\begin{equation}
    \label{eq:DP_RP_fixed}
    \mathbb{P}(\Pi_n|\alpha) = \frac{\alpha^{K_n}}{\prod_{r=0}^{n-1}(\alpha+r)}\prod_{k=1}^{K_n}(n_k-1)!,
\end{equation}
where $n_k$, for $k\in\{1,\dots,K_n\}$, denotes the number of elements in the $k$th cluster so that $\sum_{k=1}^{K_n}n_k = n$. It is clear, then, that $\alpha$ and $n$ completely determine the probability of $\Pi_n$. In fact, given a fixed $\alpha$, the prior mean of the number of clusters is given by $\mathbb{E}[K_n|\alpha] = \sum_{i=1}^{n}\frac{\alpha}{\alpha+i-1}$. Thus, larger values of $\alpha$ and $n$ imply a larger prior expected number of clusters. 
\par
On the other hand, \cite{gnedin2006exchangeable} showed that, under a random $\alpha$ with prior density $p(\alpha)$, the marginal probability of $\Pi_n$ would be given by
\begin{equation*}
    \mathbb{P}(\Pi_n) = \int_{\mathbb{R}_{+}}\frac{\alpha^{K_n}}{\prod_{r=0}^{n-1}(\alpha+r)}p(\alpha)\text{d}\alpha\prod_{k=1}^{K_n}(n_k-1)!.
\end{equation*}
When $p(\alpha)$ is the density function of a gamma-distributed random variable, there is no analytical form for $\mathbb{P}(\Pi_n)$. This is a major drawback, as practitioners cannot transparently and straightforwardly specify a prior on $\alpha$ that represents their beliefs about their scientific problems at hand. 
\par
To overcome this, \cite{zito2023bayesian} recently introduced the Stirling-gamma distribution as a prior for $\alpha$, which has the following appealing properties: (a) it leads to an analytically tractable expression for $\mathbb{P}(\Pi_n)$, (b) it leads to an approximate negative binomial prior on the number of clusters, and (c) it is conjugate to the random partition induced by a DP, so it is computationally very efficient. What follows is a brief review of the Stirling-gamma distribution and its related process. For additional details, please refer to the authors' work and the references therein.
\begin{definition}[The Stirling-gamma distribution \citep{zito2023bayesian}]
A positive random variable $\alpha$ is said to follow a Stirling-gamma distribution with parameters $a,b>0$ and $m\in\mathbb{N}$, satisfying $1<a/b<m$, denoted as $\alpha\sim\textnormal{Sg}(a,b,m)$, if its density function is given by
\begin{equation}
    \label{eq:stirling_gamma_density}
    p(\alpha) = \left[\int_{\mathbb{R}_{+}}\frac{\alpha^{a-1}}{\left(\prod_{r=0}^{m-1}(\alpha+r)\right)^{b}}\textnormal{d}\alpha\right]^{-1}\frac{\alpha^{a-1}}{\left(\prod_{r=0}^{m-1}(\alpha+r)\right)^{b}}.
\end{equation}
\end{definition}
\par
Interestingly, if $G|\alpha\sim\textnormal{DP}(\alpha,G_0)$ with $\alpha\sim\textnormal{Sg}(a,b,m)$, \cite{zito2023bayesian} showed that, among many other fascinating properties of the Stirling-gamma distribution, for a sufficiently large $m$, the prior number of clusters induced by $G$, denoted as $K_m$, would have mean $\mathbb{E}[K_m]=\frac{a}{b}$ and variance $\mathbb{V}\textnormal{ar}(K_m)\approx\frac{b+1}{b}\left(\frac{a}{b}-1\right)$. Moreover, the Stirling-gamma distribution is also conjugate to the random partition induced by a DP. In other words, if $\Pi_n|\alpha$ is distributed as in \eqref{eq:DP_RP_fixed} and $\alpha\sim\textnormal{Sg}(a,b,n)$, then $\alpha|\Pi_n\sim\textnormal{Sg}(a+K_n,b+1,n)$.
\par
The above results have notable implications. Firstly, practitioners can transparently and straightforwardly specify a prior on $\alpha$ that reflects their beliefs about their individual scientific problems at hand. For instance, letting $\alpha\sim\textnormal{Sg}(1, 0.2, n)$ implies that $\mathbb{E}[K_n]\approx5$, meaning that \textit{a-priori}, the expected number of clusters would be around five. Similarly, letting $\alpha\sim\textnormal{Sg}(10,2,n)$ also implies that $\mathbb{E}[K_n]\approx5$. The latter prior, however, is more informative than the former. As such, practitioners can elicit an informative prior on the number of clusters via $a$ and $b$, while letting $m=n$ with $n\rightarrow\infty$. Secondly, due to the conjugacy of the Stirling-gamma distribution, practitioners can efficiently perform posterior inference through a Gibbs sampling scheme. Motivated by these benefits, the remainder of this article focuses on extending the autoregressive logistic-beta DPM for dynamic clustering, from Section \ref{sec:LB_DP}, with a Stirling-gamma prior, in order to identify dynamic clusters of air-quality monitoring stations exhibiting similar FSP levels across continental Chile.

\section{The autoregressive logistic-beta Stirling-gamma process for dynamic clustering}
\label{sec:AR_SGP}
\subsection{Model formulation}
\label{subsec:the_model}
As in \cite{DeIorio2023_AR_DPMs}, we consider a sequence of time-dependent discrete random probability measures, $\{G_t\}_{t=1}^T$, such that $\{G_t\}_{t=1}^T$ has an AR structure and, for any $t\in\{1,\dots,T\}$, $G_t \sim \textnormal{DP}(\alpha_t, G_0)$. In our case, however, we construct each $G_t$ as
\begin{equation*}
    G_t(\cdot) = \sum_{k\in\mathbb{N}}\left(\textnormal{logit}^{-1}(\epsilon_k(t))\prod_{l<k}\left[1-\textnormal{logit}^{-1}(\epsilon_l(t))\right]\right)\delta_{\zeta_{k}}(\cdot), \qquad \{\zeta_{k}\}_{k\in\mathbb{N}} \overset{\text{iid}}{\sim} G_0,
\end{equation*}
and for each $\{\epsilon_k(t)\}_{t=1}^T$ we consider a generalization of the multivariate logistic-beta distribution from \cite{logistic_beta_2025}. More precisely, we let the density of $\boldsymbol{\epsilon}_k = (\epsilon_k(1),\dots, \epsilon_k(T))'$, for $k\in\mathbb{N}$, be given by
\begin{equation*}
    p(\boldsymbol{\epsilon}_k) = \int_{\mathbb{R}_+} \dots \int_{\mathbb{R}_+} \phi_{T}(\boldsymbol{\epsilon}_k;\, \frac{1}{2}\boldsymbol{\Lambda}_k(\mathbf{1}_T - \boldsymbol{\alpha}), \boldsymbol{\Lambda}_k^{1/2}\boldsymbol{\Psi}\boldsymbol{\Lambda}_k^{1/2}) \prod_{t=1}^{T}\pi_{\textnormal{Polya}}(\lambda_{kt};\,1,\alpha_t) \textnormal{d}\lambda_{k1} \dots \textnormal{d}\lambda_{kT},
\end{equation*}
where $\boldsymbol{\alpha} = (\alpha_1, \dots, \alpha_T)'\in\mathbb{R}^T$ and $\boldsymbol{\Lambda}_k=\textnormal{diag}(\lambda_{k1},\dots,\lambda_{kT})\in\mathbb{R}^{T \times T}$, with $\phi_d(\cdot;\boldsymbol{\mu},\boldsymbol{\Sigma})$ denoting the density function of a $d$-dimensional multivariate normally-distributed random variable with mean vector $\boldsymbol{\mu}\in\mathbb{R}^{d}$ and covariance matrix $\boldsymbol{\Sigma}\in\mathbb{R}^{d\times d}$. Lastly, to allow for an AR structure in $\{G_t\}_{t=1}^T$, via $\{\boldsymbol{\epsilon}_k\}_{k\in\mathbb{N}}$, we let $\boldsymbol{\Psi}$ be an AR(1) correlation matrix of the form $\boldsymbol{\Psi}_{t, t'} = \psi^{|t-t'|}$.
\par
Following \cite{logistic_beta_2025}, the conditional marginal distribution of the $t$th entry of $\boldsymbol{\epsilon}_k$ would then be given by
\begin{equation*}
    \epsilon_{k}(t)|\lambda_{kt} \sim N(0.5\lambda_{kt}(1 - \alpha_t),\, \lambda_{kt}), \qquad \lambda_{kt} \sim \textnormal{Polya}(1, \alpha_t). 
\end{equation*}
Thus, by integrating out $\lambda_{kt}$, $\epsilon_{k}(t)$ would follow a logistic-beta distribution with parameters $1$ and $\alpha_t$ so that $\textnormal{logit}^{-1}(\epsilon_{k}(t))\sim\textnormal{Beta}(1, \alpha_t)$, defining a proper DP prior over $G_t$. Our generalization, though, allows for non-identical beta marginals across time points, namely, $\textnormal{logit}^{-1}(\epsilon_{k}(t)) \sim \textnormal{Beta}(1,\alpha_t)$, rather than the ``one-size fits all" approach of traditional logistic-beta dependent DPs, where a single concentration parameter $\alpha$ governs the entire stick-breaking mechanism at every time point, i.e., for all $t\in\mathbb{N}$,  $\textnormal{logit}^{-1}(\epsilon_{k}(t)) \sim \textnormal{Beta}(1,\alpha)$. By allowing for different $\alpha$-values, we can capture more easily idiosyncratic clustering structures across time points, rather than shrinking all time-specific partitions toward a common clustering pattern. In fact, in the Supplementary Materials, we show that in our real-world data analyses, different months and seasons are associated with distinct $\alpha$-values, providing empirical support for this added flexibility.
\par
Lastly, to facilitate a transparent and straightforward prior elicitation, we follow \cite{zito2023bayesian} and let
\begin{equation*}
    \alpha_1,\dots, \alpha_T \overset{\text{iid}}{\sim} \textnormal{Sg}(a, b, n).
\end{equation*}
As such, we say that $\{G_t\}_{t=1}^T$ follows an \textit{autoregressive logistic-beta Stirling-gamma} (AR-LB-SG) process, denoted by $\{G_t\}_{t=1}^T\sim\textnormal{AR-LB-SG}(\psi, G_0, a, b, n)$, which is computationally more efficient than its copula-based counterpart. 
\par
In many applications, however, like when modeling the behavior of air-pollutants, the discrete nature of each $G_t$ might be too restrictive. Following \cite{Antoniak_DPM} and \cite{Lo_DPMs1984}, a natural solution is then to incorporate our AR-LB-SG process within a hierarchical model. More formally, let $\{(Y_{1t},\dots,Y_{nt})\}_{t=1}^{T}$ be our dependent sequence of observables, where each $Y_{it}$ denotes the FSP concentration level from station $i$ at time $t$ 
such that
\begin{equation}
    \label{eq:Normal_parametric_kernel}
    Y_{it}|\mu_{it},\sigma_{it}^{2}\overset{\text{ind}}{\sim}N(\theta_{it} + \mathbf{x}_{it}'\boldsymbol{\beta} + \gamma_i,\sigma_{it}^{2}).
\end{equation}
where $\boldsymbol{\gamma}=(\gamma_1,\dots,\gamma_n)'\in\mathbb{R}^{n}$ is a vector of location-specific random effects, $\mathbf{x}_{it}\in\mathbb{R}^{p}$ is a vector of covariates associated with station $i$ at time $t$, and $\boldsymbol{\beta}=(\beta_1,\dots,\beta_p)'\in\mathbb{R}^{p}$ is its corresponding vector of coefficients with $\boldsymbol{\beta} | \rho^2 \sim N_p(\mathbf{0}_p,\,\rho^2\mathbf{I}_p)$. To introduce spatial dependence among the $n$ stations, we assume a spatially correlated prior of the form
\begin{equation}
    \label{eq:spatial_prior}
    \boldsymbol{\gamma}|\mathbf{K},\tau^{2},\varphi\sim N_n\left(\mathbf{0}_n,\, \mathbf{K}\right),
\end{equation}
where $\mathbf{K}$ is a squared exponential kernel function of size $n\times n$ such that $(\mathbf{K})_{i,i'}= \tau^2\exp\left\{-(d_{i,i'})^2/(2\varphi^2)\right\}$, $d_{i,i'}$ is the geographical distance between stations $i$ and $i'$, and $\varphi$ and $\tau^{2}$ control the range of the spatial correlation and the amount of total variation between the location-specific random effects, respectively.
\par
Lastly, to allow for dynamic clustering, we let, for any $t\in\{1,\dots,T\}$,
\begin{equation}
    \label{eq:ARSGP_hyperprior}
    \begin{split}
        (\theta_{1t},\sigma_{1t}^2),\dots,(\theta_{nt},\sigma_{nt}^2)|G_t\overset{\text{iid}}{\sim}G_t, \quad \\
        \{G_t\}_{t\in\mathbb{T}}\sim\textnormal{AR-LB-SG}(\psi,G_0,a,b,n). \\
    \end{split}
\end{equation}
\par
To complete our model specification, we let $\psi \sim \textnormal{Unif}(-1,1)$, $\varphi\sim\textnormal{Ga}(a_{\varphi},b_{\varphi})$, $\rho^2\sim\textnormal{IG}(a_{\rho},b_{\rho})$, and $\tau^2 \sim \textnormal{IG}(a_{\tau},b_{\tau})$, where $\textnormal{Unif}(-1,1)$ denotes the uniform distribution on $(-1,1)$, $\textnormal{Ga}(a_{\varphi},b_{\varphi})$ denotes the gamma distribution with mean $\frac{a_{\varphi}}{b_{\varphi}}$, and $\textnormal{IG}(a_x,b_x)$ denotes the inverse-gamma distribution with mean $\frac{b_x}{a_x-1}$. Additionally, we set $G_0$ so that
\begin{equation}
    \label{eq:ARSGP_base_measure}
    G_0(\textnormal{d}(\theta,\sigma^2))\propto\sigma_0^{-1}\left(\sigma^2\right)^{-a_0-1}\exp\left\{-\left[ \frac{(\theta - \theta_0)^2}{2\sigma_0^2}+\frac{b_0}{\sigma^2}\right]\right\}, \quad (\theta,\sigma^{2})\in\mathbb{R}\times\mathbb{R}_{+},
\end{equation}
where $\theta_0\in\mathbb{R}$ and $a_0,\,b_0,\,\sigma^2_0>0$ are user-specified values.
\par
In the above model, the discrete nature of each $G_t$, for $t\in\{1,\dots,T\}$, induces a clustering of the observations $Y_{1t}, \dots, Y_{nt}$. More formally, at each discrete time $t$, there will be $K_{n,t}\in[1,n]$ distinct pairs among $\{ (\theta_{1t},\sigma_{1t}^2),\dots,(\theta_{nt},\sigma_{nt}^2)\}$, separating the $n$ observations into $K_{n,t}$ clusters \citep{muller2015bayesian}. In other words, we say that, at time $t$, $Y_{it}$ and $Y_{i't}$, for $i\neq i'$, are clustered together if an only if $(\theta_{it},\sigma_{it}^2) = (\theta_{i't},\sigma_{i't}^2)$. Thus, let us define $\Pi_{n,t}$ as the random partition induced by $G_t$ at any discrete time $t$.
\par
It is clear, then, that the AR-LB-SG process allows for dynamic clustering (i.e., it allows for membership allocations and number of clusters to evolve over time) in a very similar fashion to the copula-based autoregressive DPM from \cite{DeIorio2023_AR_DPMs}. Our proposed model, though, inherits all the computational benefits from the logistic-beta dependent DP, as well as the transparency and interpretability of the Stirling-gamma prior on the induced number of clusters, making the AR-LB-SG process suitable in many applied settings. 
\par
An alternative approach for dynamic clustering is the \textit{dependent random partition model} from \cite{page2022dependent}, in which the authors directly model a sequence of random partitions indexed by discrete time, rather than modeling a sequence of discrete random measures that will induce a sequence of (latent) random partitions. Through various simulation studies and real-world data analyses, \cite{page2022dependent} demonstrate the remarkably strong statistical and computational performance of their proposed method. Thus, in subsequent sections, the dependent random partition model is employed as a benchmark.

\subsection{Posterior inference}
\label{subseq:posterior_inf}
Since the posterior distribution implied by model \eqref{eq:Normal_parametric_kernel}--\eqref{eq:ARSGP_base_measure} is not of a known form, we generate random draws from such a distribution via Markov chain Monte Carlo (MCMC). We will now present the full conditional distributions of the time-dependent stick-breaking weights, $w_{tk}$, and the concentration parameters, $\alpha_t$, as these are central to our method. Implementation details of the whole MCMC algorithm are included in the Supplementary Materials.

\subsubsection{Full conditional distribution of the cluster membership indicators}
Let us introduce latent cluster membership indicators, $\{s_{1t},\dots,s_{nt}\}_{t=1}^{T}$, such that $s_{it}=k$ if $Y_{it}$ is allocated to cluster $k$ at time $t$. Similarly, let $(\theta_{kt},\sigma^2_{kt})$ be the unique parameters in cluster $k$ at time $t$. Following \cite{logistic_beta_2025} and \cite{DeIorio2023_AR_DPMs}, we also truncate $G_{t}(\cdot)=\sum_{k\in\mathbb{N}}w_{tk}\delta_{\zeta_{k}}(\cdot)$ to $H$ terms so that $\nu_{tH}\equiv1$, where $H$ is chosen to be sufficiently large \citep{Ishwaran_James_2001_stick, Gelfand_Kottas_2002_Dirichlet}. Additional details regarding the choice of $H$ are given in the Supplementary Materials.
\par
Moreover, since the membership indicators at different time points are conditionally independent given $\{w_{t1},\dots,w_{tH}\}_{t=1}^{T}$, the full conditional distribution of $\{s_{1t},\dots,s_{nt}\}_{t=1}^{T}$ can be factorized as the product of the individual full conditional distributions of each membership indicator \citep{DeIorio2023_AR_DPMs}. As such, we can sample each $s_{it}$, for $i\in\{1,\dots,n\}$ and $t\in\{1,\dots,T\}$, from
\begin{equation*}
    \mathbb{P}(s_{it} = k|\Omega_{-s_{it}})\propto w_{tk}\,\phi(y_{it};\,\theta_{kt} + \mathbf{x}_{it}'\boldsymbol{\beta} + \gamma_i,\sigma_{kt}^2),\qquad k\in\{1,\dots,H\},
\end{equation*}
where $\Omega_{-s_{it}}$ denotes the set of all other model parameters and data, excluding $s_{it}$.
\subsubsection{Full conditional distribution of the stick-breaking weights}

Let us now introduce the index set $\mathcal{I}_{k} = \{(i, t):s_{it}>k-1\}$ and let $\mathcal{Z}_{k}=\{z_{ikt} = \mathbbm{1}\{s_{it} = k\}\in\{0, 1\}: (i,t) \in\mathcal{I}_{k}\}$ be a set of latent binary variables such that $z_{ikt} = 1$ if $s_{it}=k$ and $z_{ikt} = 0$ if $s_{it}>k$. Additionally, let $m_{kt}=\sum_{i}\mathbbm{1}\{s_{it}>k-1\}$ and $r_{kt} = \sum_{i}z_{ikt}$. It is clear that, conditional on $\epsilon_k(t)$, $r_{kt}$ follows a binomial distribution with $m_{kt}$ trials. As such, we can sample from the full conditional distribution of the time-dependent stick-breaking weights---via $\epsilon_{k}(t)$---through a modified version of the data-augmentation Metropolis-within-Gibbs sampling scheme from \cite{logistic_beta_2025}. More precisely, for $k\in\{1,\dots,H-1\}$, such an algorithm would proceed as follows:
\begin{enumerate}
    \item Let $\mathbb{T}=\{1,\dots,T\}$ and let $\mathbb{T}_k = \{t: m_{kt} > 0 \}\subseteq\mathbb{T}$ be the subset of time points---with cardinality $\textnormal{card}({\mathbb{T}}_k)={T}_k$---for which $m_{kt} > 0$. Then, for all $t\in{\mathbb{T}}_k$, sample
    \begin{equation*}
        \xi_{kt}|\epsilon_{k}(t) \overset{\text{ind}}{\sim} \textnormal{PG}(m_{kt},\epsilon_{k}(t)),
    \end{equation*}
    where $\textnormal{PG}(a_\xi, b_\xi)$ denotes the P\'olya-gamma distribution with parameters $a_\xi$ and $b_\xi$ as in \cite{Polson_ploya_gamma}. For all $t\notin\mathbb{T}_k$, on the other hand, set $\xi_{kt} = 0$.
    \item Let $\Tilde{\mathbf{m}}_k = (m_{kt} : t\in\mathbb{T}_k)'\in\mathbb{R}^{T_k}$, $\Tilde{\mathbf{r}}_k = (r_{kt}:t\in\mathbb{T}_k)'\in\mathbb{R}^{T_k}$, $\Tilde{\boldsymbol{\alpha}}_k = (\alpha_t : t\in\mathbb{T}_k)'\in\mathbb{R}^{T_k}$, and $\Tilde{\boldsymbol{\lambda}}_k = (\lambda_{kt} : t\in\mathbb{T}_k)'\in\mathbb{R}^{T_k}$ be $T_k$-dimensional vectors containing the values of $m_{kt}$, $r_{kt}$, $\alpha_t$, and $\lambda_{kt}$ for which $t\in{\mathbb{T}}_k$. Similarly, let $\Tilde{\boldsymbol{\Xi}}_k=\textnormal{diag}(\xi_{kt} : t\in{\mathbb{T}}_k)\in\mathbb{R}^{{T}_k\times {T}_k}$, $\Tilde{\boldsymbol{\Lambda}}_k = \textnormal{diag}(\Tilde{\boldsymbol{\lambda}}_k)\in\mathbb{R}^{T_k\times T_k}$, and let $\Tilde{\boldsymbol{\Psi}}_k$ be a ${T}_k\times {T}_k$ matrix with entries $(\Tilde{\boldsymbol{\Psi}}_k)_{t,t'}=\psi^{|t-t'|}$ such that $t,t'\in{\mathbb{T}}_k$. Then, sample $\Tilde{\boldsymbol{\lambda}}_k$ from
    \begin{equation}
        \label{eq:sample_lambdas}
        \begin{split}
            p(\Tilde{\boldsymbol{\lambda}}_k|\Omega_{-\boldsymbol{\lambda}_k})  \propto  \phi_{T_k}(\Tilde{\boldsymbol{\Xi}}_k^{-1}(\Tilde{\mathbf{r}}_k - \frac{1}{2}\Tilde{\mathbf{m}}_k);\, \Tilde{\boldsymbol{a}}_k, \Tilde{\mathbf{B}}_k ) \prod_{t\in\mathbb{T}_k} \pi_{\textnormal{Polya}}(\lambda_{kt};\,1,\alpha_t),
        \end{split}
    \end{equation}
    where $\Tilde{\boldsymbol{a}}_k = \frac{1}{2}\Tilde{\boldsymbol{\Lambda}}_k(\mathbf{1}_{T_k} - \Tilde{\boldsymbol{\alpha}}_k)$ and $\Tilde{\mathbf{B}}_k = \Tilde{\boldsymbol{\Lambda}}_k^{1/2}\Tilde{\boldsymbol{\Psi}}_k\Tilde{\boldsymbol{\Lambda}}_k^{1/2} + \Tilde{\boldsymbol{\Xi}}_k^{-1}$. Additionally, for all $t\notin\mathbb{T}_k$, sample $\{\lambda_{kt}\}_{t\notin\mathbb{T}_k} \overset{\text{ind}}{\sim} \textnormal{Polya}(1, \alpha_t)$.
    \label{step:sample_lambda}
    \item Sample $\boldsymbol{\epsilon}_k=(\epsilon_k(1),\dots,\epsilon_k(T))'$ from
    \begin{equation*}
        \boldsymbol{\epsilon}_k|\Omega_{-\boldsymbol{\epsilon}_k}\sim N_{T}\left(\hat{\boldsymbol{\Xi}}_k^{-1}\hat{\boldsymbol{e}}_k,\, \hat{\boldsymbol{\Xi}}_k^{-1} \right),
    \end{equation*}
    where $\hat{\boldsymbol{\Xi}}_k = \boldsymbol{\Xi}_{k} + (\boldsymbol{\Lambda}_k^{1/2}\boldsymbol{\Psi}\boldsymbol{\Lambda}_k^{1/2})^{-1}$ and $\hat{\boldsymbol{e}}_k = (\mathbf{r}_k - \frac{1}{2}\mathbf{m}_k) + (\boldsymbol{\Lambda}_k^{1/2}\boldsymbol{\Psi}\boldsymbol{\Lambda}_k^{1/2})^{-1}[\frac{1}{2}\boldsymbol{\Lambda}_k(\mathbf{1}_T - \boldsymbol{\alpha})]$. Here, $\mathbf{r}_k = (r_{k1}, \dots, r_{kT})'$, $\mathbf{m}_k = (m_{k1}, \dots, m_{kT})'$, $\boldsymbol{\alpha} = (\alpha_1, \dots, \alpha_T)'$, $\boldsymbol{\Xi}_{k}  = \textnormal{diag}(\xi_{k1}, \dots, \xi_{kT})$, $\boldsymbol{\Lambda}_k = \textnormal{diag}(\lambda_{k1},\dots, \lambda_{kT})$, and $(\boldsymbol{\Psi})_{t,t'}=\psi^{|t-t'|}$ such that $t,t'\in\mathbb{T}$.
    \label{step:sample_epsilon}
    \item Set
    \begin{equation*}
        w_{tk}=\textnormal{logit}^{-1}(\epsilon_{k}(t))\prod_{l<k}\left[1-\textnormal{logit}^{-1}(\epsilon_{l}(t))\right],
    \end{equation*}
    with $\textnormal{logit}^{-1}(\epsilon_{H}(t))\equiv1$.
\end{enumerate}

\par

Note, however, that sampling $\Tilde{\boldsymbol{\lambda}}_k$ from \eqref{eq:sample_lambdas} is still a nontrivial task as it requires the evaluation of $T_k$ P\'olya density functions. Fortunately, though, the following result holds, allowing us to bypass the direct evaluation of $\pi_{\textnormal{Polya}}$.
\begin{proposition}
\label{prop:MH_ratio}
For a suitable chosen set of pairs $\{(a_{\lambda_t}', b_{\lambda_t}') : t\in\mathbb{T}_k\}$ such that $a_{\lambda_t}' + b_{\lambda_t}' = 1 + \alpha_t$, the Metropolis-Hastings acceptance ratio for the proposals $\Tilde{\boldsymbol{\lambda}}_k^* = (\lambda_{kt}^* : t\in\mathbb{T}_k)' \overset{\text{ind}}{\sim} \textnormal{Polya}(a_{\lambda_t}', b_{\lambda_t}')$ becomes $\textnormal{min}\{1, r_{\boldsymbol{\lambda}_k}\}$, where
\begin{equation*}
    r_{\boldsymbol{\lambda}_k} = \exp \left\{\mathcal{L}_k(\Tilde{\boldsymbol{\lambda}}_k^*) - \mathcal{L}_k(\Tilde{\boldsymbol{\lambda}}_k) + \frac{1}{2}\sum_{t\in\mathbb{T}_k}(\alpha_t - a_{\lambda_t}'b_{\lambda_t}')(\lambda_{kt} - \lambda_{kt}^{*}) \right\},
\end{equation*}
with 
\begin{equation*}
    \mathcal{L}_k(\Tilde{\boldsymbol{\lambda}}_k) = \log \phi_{T_k}(\Tilde{\boldsymbol{\Xi}}_k^{-1}(\Tilde{\mathbf{r}}_k - \frac{1}{2}\Tilde{\mathbf{m}}_k);\, \frac{1}{2}\Tilde{\boldsymbol{\Lambda}}_{\Tilde{\boldsymbol{\lambda}}_k}(\mathbf{1}_{T_k} - \Tilde{\boldsymbol{\alpha}}_k), \Tilde{\boldsymbol{\Lambda}}_{\Tilde{\boldsymbol{\lambda}}_k}^{1/2}\Tilde{\boldsymbol{\Psi}}_k\Tilde{\boldsymbol{\Lambda}}_{\Tilde{\boldsymbol{\lambda}}_k}^{1/2} + \Tilde{\boldsymbol{\Xi}}_k^{-1}),
\end{equation*}
such that, for any vector $\boldsymbol{\ell} = (\ell_t : t\in\mathbb{T}_k)'$, $\boldsymbol{\ell} \mapsto \Tilde{\boldsymbol{\Lambda}}_{\boldsymbol{\ell}} = \textnormal{diag}(\ell_t : t\in\mathbb{T}_k)$.
\end{proposition}

\par

Details on the derivation of Proposition \ref{prop:MH_ratio} are presented in the Supplementary Materials. Moreover, to set $\{(a_{\lambda_t}', b_{\lambda_t}') : t\in\mathbb{T}_k\}$ in an adaptive fashion, we follow \cite{logistic_beta_2025} and employ a moment matching method with a running average of $\Tilde{\boldsymbol{\lambda}}_k$.

\subsubsection{Full conditional distributions of \texorpdfstring{$\alpha_1,\dots,\alpha_T$}{alpha\_1, ..., alpha\_T}}
Since, for any $t\in\{1,\dots,T\}$, $G_t|\alpha_t\sim\textnormal{DP}(\alpha_t, G_0)$, then $\{\Pi_{n,t}|\alpha_t\}$ is going to be distributed as in \eqref{eq:DP_RP_fixed}, where $\Pi_{n,t}$ denotes the random partition induced by $G_t$ at any discrete time $t$. As such, by combining \eqref{eq:DP_RP_fixed} and \eqref{eq:stirling_gamma_density}, we have that the full conditional distribution of $\alpha_t$, for $t\in\{1,\dots,T\}$, is given by
\begin{equation}
    \begin{split}
        p(\alpha_t|\Pi_{n,t}) & \propto p(\alpha_t)\, \mathbb{P}(\Pi_{n,t}|\alpha_t) \propto \frac{\alpha_t^{(a+K_{n,t})-1}}{\left(\prod_{r=0}^{n-1}(\alpha_t + r)\right)^{b+1}}.
    \end{split}
\end{equation}
Thus, $\{\alpha_t|\Pi_{n,t}\}_{t=1}^{T} \overset{\text{ind}}{\sim} \textnormal{Sg}(a + K_{n,t}, b+1, n)$, where $K_{n,t}$ denotes the total number of clusters at each discrete time $t$ (see also \cite{zito2023bayesian}). 

\subsection{Stirling-gamma versus gamma priors}
Lastly, we would like to compare the Stirling-gamma and gamma priors on $\alpha_t$ within the context of autoregressive DP mixtures. More precisely, let $\{G_t\}_{t=1}^T$ be a time-dependent sequence of discrete random probability measures with an AR structure similar to the one in \eqref{eq:Normal_parametric_kernel}--\eqref{eq:ARSGP_base_measure} so that, for any $t\in\{1,\dots,T\}$, $G_t|\alpha_t \sim\textnormal{DP}(\alpha_t,G_0)$. In this case, however, we let $\alpha_1,\dots,\alpha_T \overset{\text{iid}}{\sim} \textnormal{Ga}(a, b)$ as in \cite{escobar1995bayesian}. As such, the full conditional distribution of each $\alpha_t$, for $t\in\{1\dots,T\}$, would be given by
\begin{equation}
    \begin{split}
        p(\alpha_t|\Pi_{n,t}) & \propto \exp\left(-\alpha_t b \right) \bigg( \alpha_t^{a -1 + K_{n,t}} + n\alpha_t^{a-2+K_{n,t}} \bigg) \int_0^1 x^{\alpha_t}(1-x)^{n-1}\textnormal{d}x .
    \end{split}
\end{equation}
\par
Conditional on an auxiliary beta-distributed random variable, such a full conditional distribution reduces to a two-component mixture of gamma distributions (see \cite{escobar1995bayesian}). Details on how to generate random draws from the above distribution are included in the Supplementary Materials. A detailed comparison of the clustering results obtained under a Stirling-gamma and a gamma prior is also presented in the Supplementary Materials, where we can observe minimal differences between the results under the two different priors on $\alpha_1,\dots,\alpha_T$. This is not surprising, due to the close similarities between the gamma and the Stirling-gamma distributions \citep{zito2023bayesian}. That being said, for applications involving hazardous air pollutants---as well as other high-stakes decision-making scenarios---we consider the Stirling-gamma prior to be a natural choice as it allows for a more transparent and straightforward workflow. 

\section{Numerical studies}
\label{sec:simulations}
To evaluate the finite sample performance of the AR-LB-SG process, we now conduct a series of numerical studies. In particular, we consider $n=64$ different locations across $T=60$ different time points---to mimic the setting in our real-world data analysis. Locations are randomly generated within continental Chile with the \textbf{R} packages "\texttt{rnaturalearth}" \citep{rnaturalearth} and "\texttt{sf}" \citep{sf}. Each covariate vector, $\mathbf{x}_{it}\in\mathbb{R}^5$, is generated uniformly at random on $(0,1)^5$. The vectors of coefficients and location-specific random effects are generated as $\beta_1,\dots,\beta_5\overset{\text{iid}}{\sim}N(3, 1)$ and $(\gamma_1,\dots,\gamma_n)'\sim N_{n}(3\times\mathbf{1}_n,\mathbf{K})$, with $\tau^2=2$ and $\varphi=100$, so that spatial correlation declines with distance but remains strong within 100 km. 
\par
To generate $\theta_{it}$ and $\sigma^2_{it}$, we consider two different scenarios of clustering patterns: balanced and partially unbalanced clusters. In both scenarios, we assume the existence of three clusters with $\theta_{it}\in\{6, 33, 60\}$ and $\sigma_{it}^2 \equiv1$. 
Under the balanced clusters case, we set the membership indicators so that, at time $t=1$, all data points are randomly assigned to any of the three clusters, each with equal probability. Then, at each time $t\geq 2$, each data point is allowed to ``jump'' to any of the other two clusters, each with probability $0.5\times \pi_{\textnormal{jump}}$, or remain in the current cluster with probability $1-\pi_{\textnormal{jump}}$. Under a partially unbalanced clusters case, we instead initialize the membership indicators so that, at time $t=1$, 70\% of the data points are randomly assigned to one cluster, while the remaining data points are randomly assigned to either of the two remaining clusters, each with equal probability. For $t\geq 2$, observations allocated to the biggest cluster are then allowed to ``jump'' to either one of the two remaining clusters with probability $\pi_{\textnormal{jump}}$ or remain in the same cluster with probability $1-\pi_{\textnormal{jump}}$. The cluster receiving the observations from the biggest cluster is chosen uniformly at random with probability 0.5. Observations allocated to the other two smaller clusters are also allowed to
``jump'' to any of the other clusters, each with probability $0.5\times\pi_{\textnormal{jump}}$, or remain in the current cluster with probability $1-\pi_{\textnormal{jump}}$.
This construction induces cluster allocations that are initially unbalanced, but whose expected proportions converge to the balanced allocation as $t\rightarrow\infty$. The parameter \(\pi_{\textnormal{jump}}\in(0,0.5]\) controls the similarity between partitions across consecutive time points. Letting \(\pi_{\textnormal{jump}}\rightarrow0\) implies that data points are more likely to remain in their current clusters (i.e., a stronger temporal dependence). Letting $\pi_{\textnormal{jump}}\rightarrow0.5$, on the other hand, implies a stronger shuffling of the cluster allocations between time points. Throughout these simulation exercises, we set $\pi_{\textnormal{jump}} = 0.05$. That being said, in the Supplementary Materials we also consider other values of $\pi_{\textnormal{jump}}$.

\par
We compare the performance of our proposed method against the dependent random partition model (DRPM, \cite{page2022dependent}) as implemented in the \textbf{R} package "\texttt{drpm}" with its default parameter settings. We employ the DRPM as a benchmark because it is the state-of-the-art method for dynamic clustering and it has effectively been applied to identify clusters of air-quality monitoring stations across rural Germany \citep{page2022dependent}. For our AR-LB-SG process, we set $a_{\varphi}=b_{\varphi}=0.1$, $a_{\rho}=b_{\rho}=0.1$, $a_{\tau}=b_{\tau}=0.1$, $a_0 = b_0 = 0.1$, $\theta_0=\bar{y} = (n\times T)^{-1} \sum_{i,t} y_{it}$, and $\sigma_0^2 = 2s^2 = 2\times (n\times T - 1)^{-1} \sum_{i,t}(y_{it} - \bar{y})^2$. Lastly, we set our Stirling-gamma prior hyperparameters to $a=1$ and $b=0.25$, which are the default values suggested by \cite{zito2023bayesian} and \cite{escobar1995bayesian}. We consider that this is a reasonable default prior specification but, clearly, different scientific applications may require different and more tailored prior distributions. In all cases, we run the MCMC algorithms for 20000 iterations, discard the first 10000 as \textit{burn-in}, and \textit{thin} the sequences by keeping every fifth draw. We execute all our simulations and data analyses on a personal laptop running on an Apple Silicon chip with 48 GB of memory and 16 CPU cores. We do not consider copula-based dependent DPs due to their prohibitively slow runtimes and the lack of open-source routines for fitting these models.
\par
\begin{figure}
\begin{center}
    \includegraphics[width=0.9\linewidth]{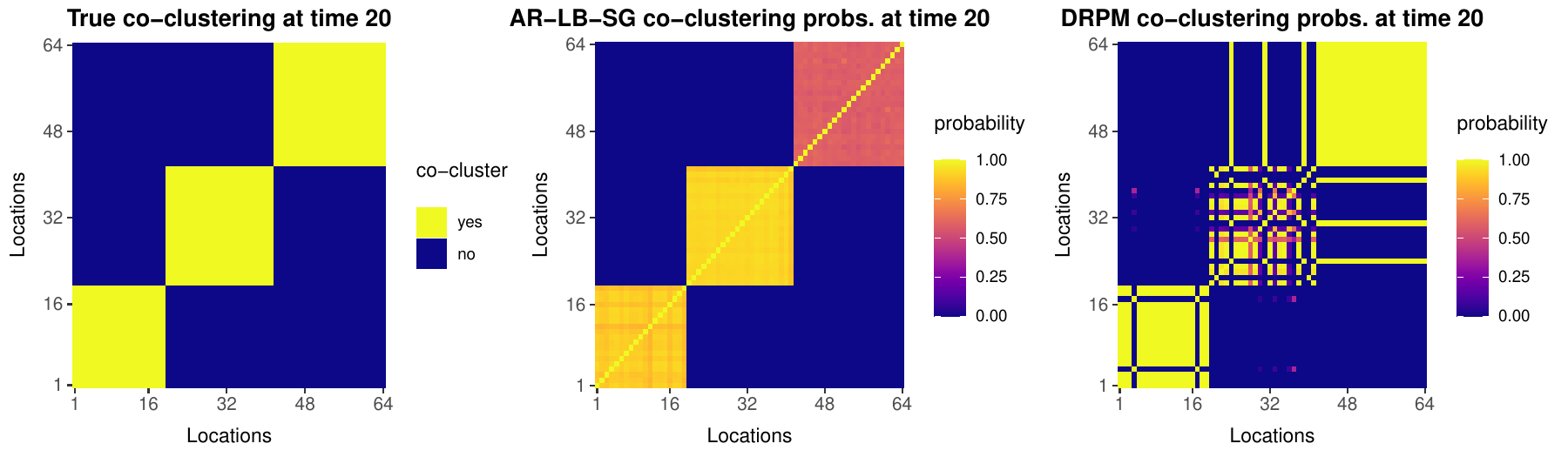} 
    
    \vspace{0.3cm}
    
    \includegraphics[width=0.9\linewidth]{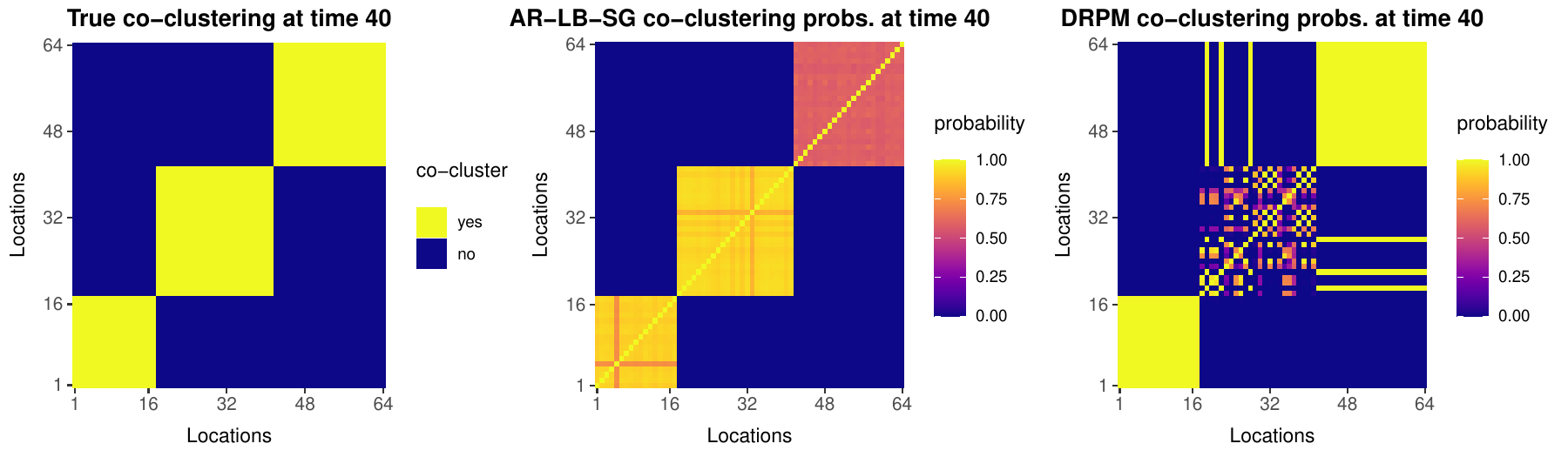} 
    
    \vspace{0.3cm}
    
    \includegraphics[width=0.9\linewidth]{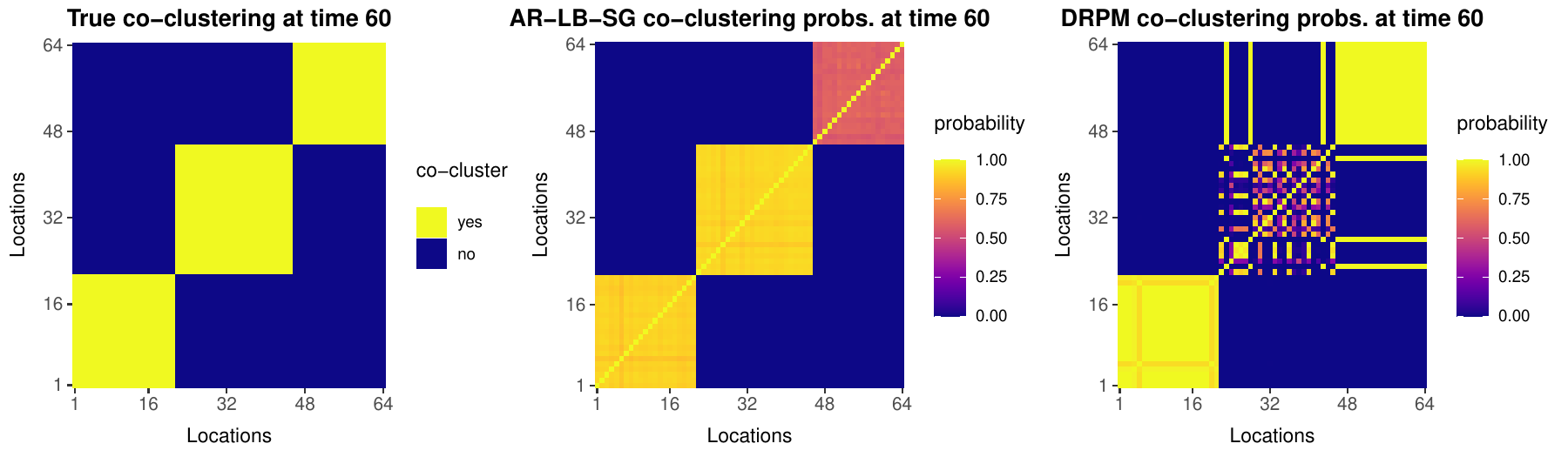}
\end{center}
\caption{True co-clustering structures and their corresponding posterior co-clustering probabilities recovered by the AR-LB-SG process and the DRPM at three different time points.}
\label{fig:coclust_sim1}
\end{figure}
\begin{figure}
\begin{center}
    \includegraphics[width=0.48\linewidth]{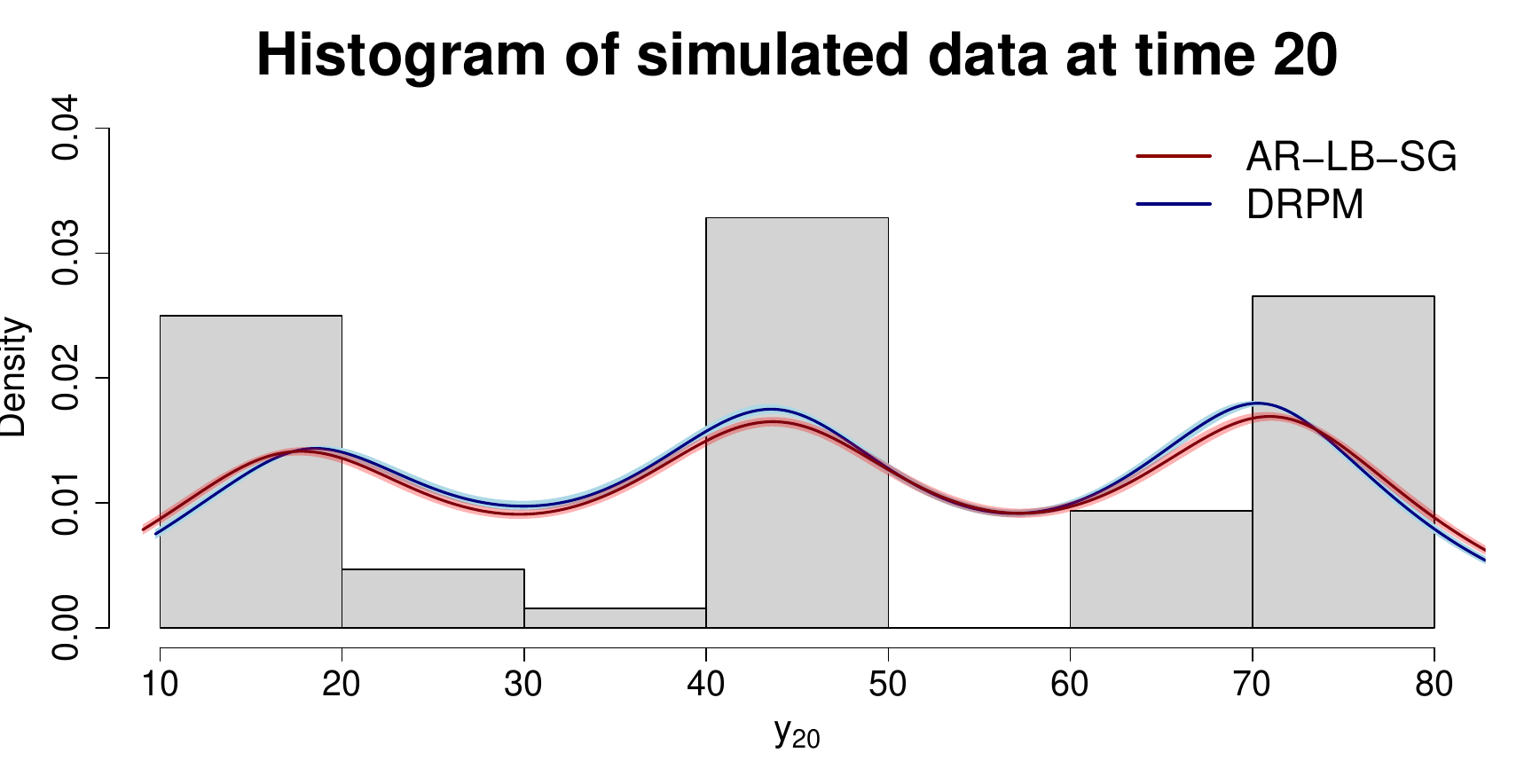}
    \includegraphics[width=0.48\linewidth]{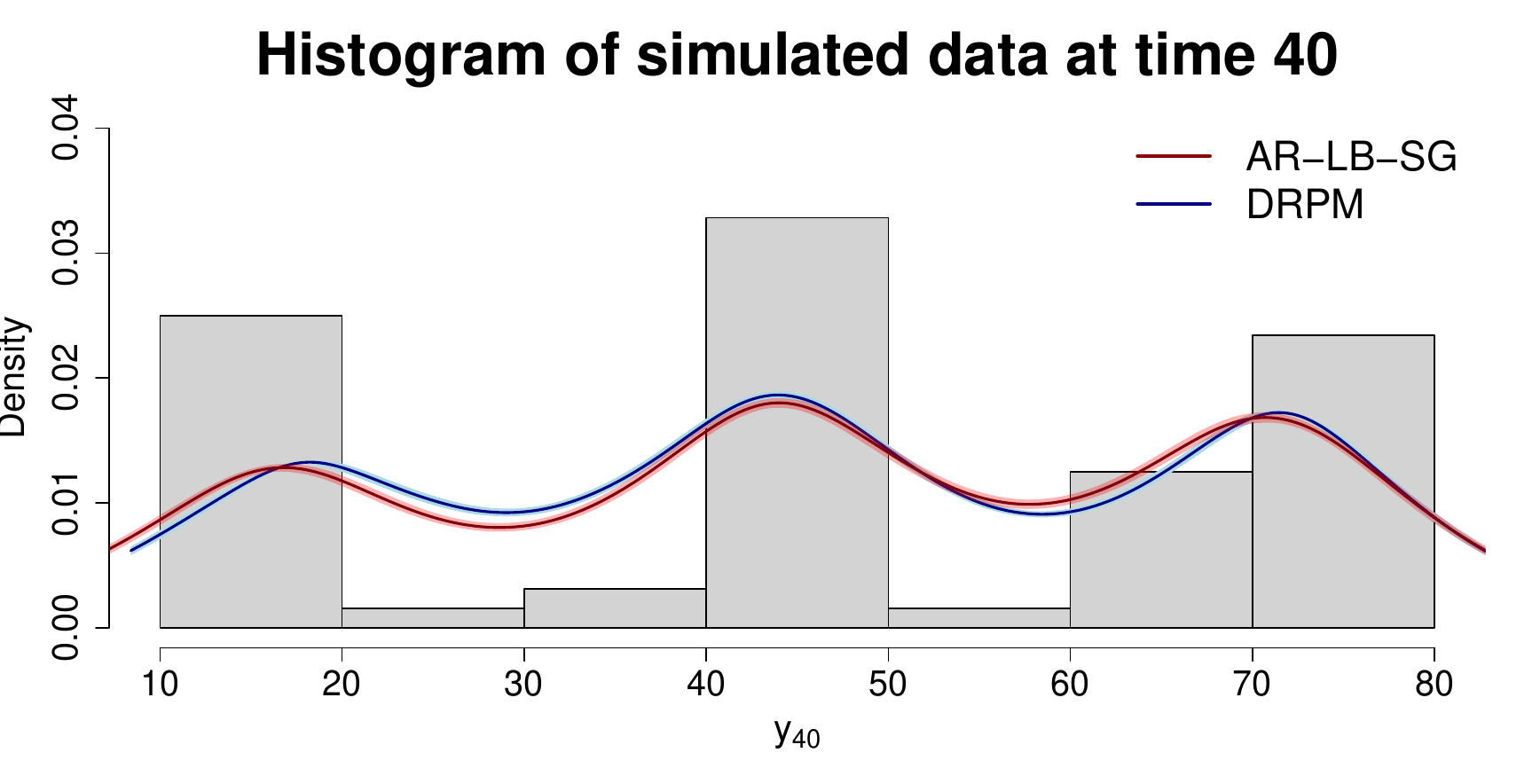}
    \includegraphics[width=0.48\linewidth]{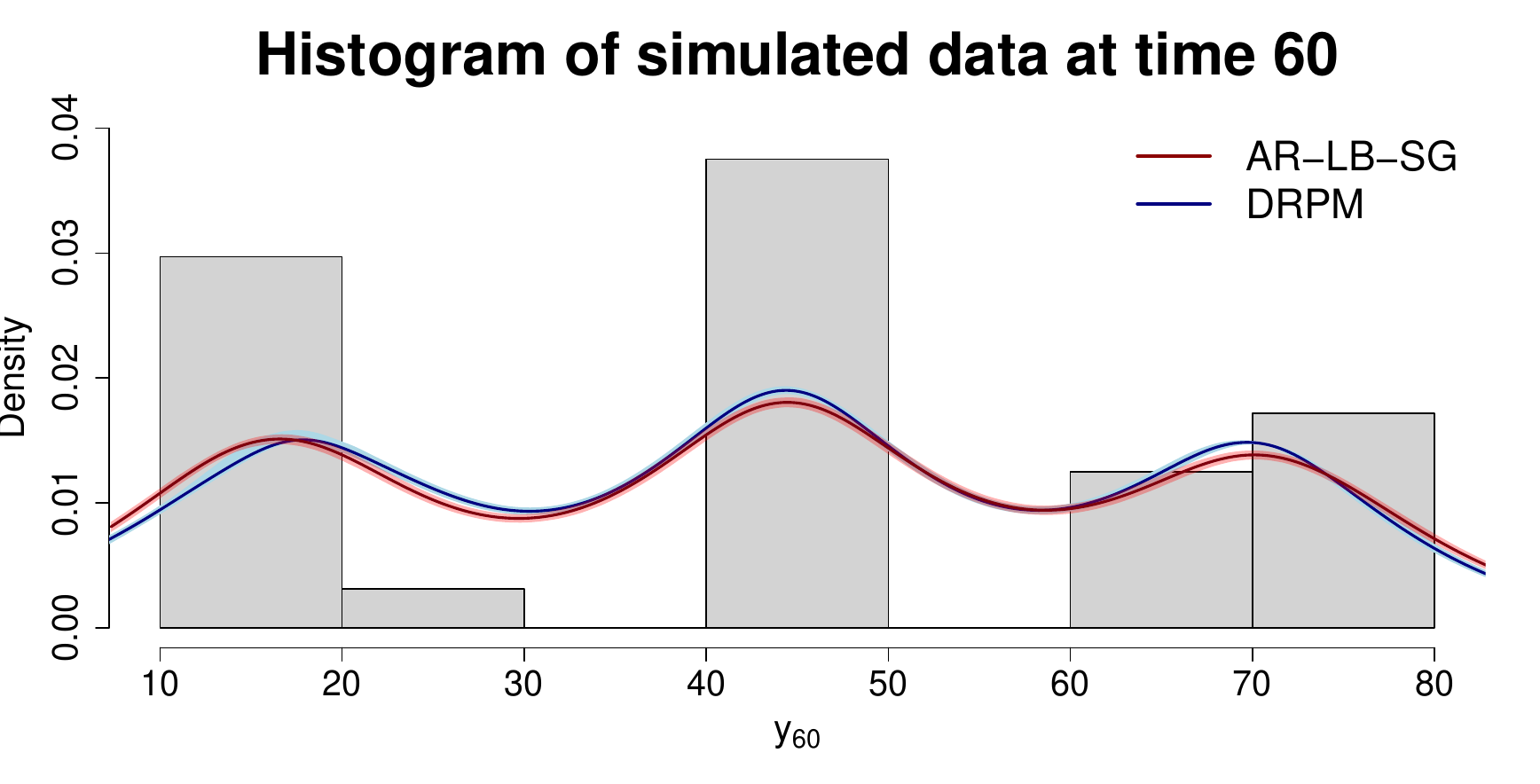}
\end{center}
\caption{Histograms of the simulated data and their corresponding posterior predictive distributions recovered by the AR-LB-SG process and the DRPM at three different time points. Tinted areas denote 95\% posterior predictive credible sets.}
\label{fig:hists_sim1}
\end{figure}
\begin{figure*}
\begin{center}
\includegraphics[width=0.9\linewidth]{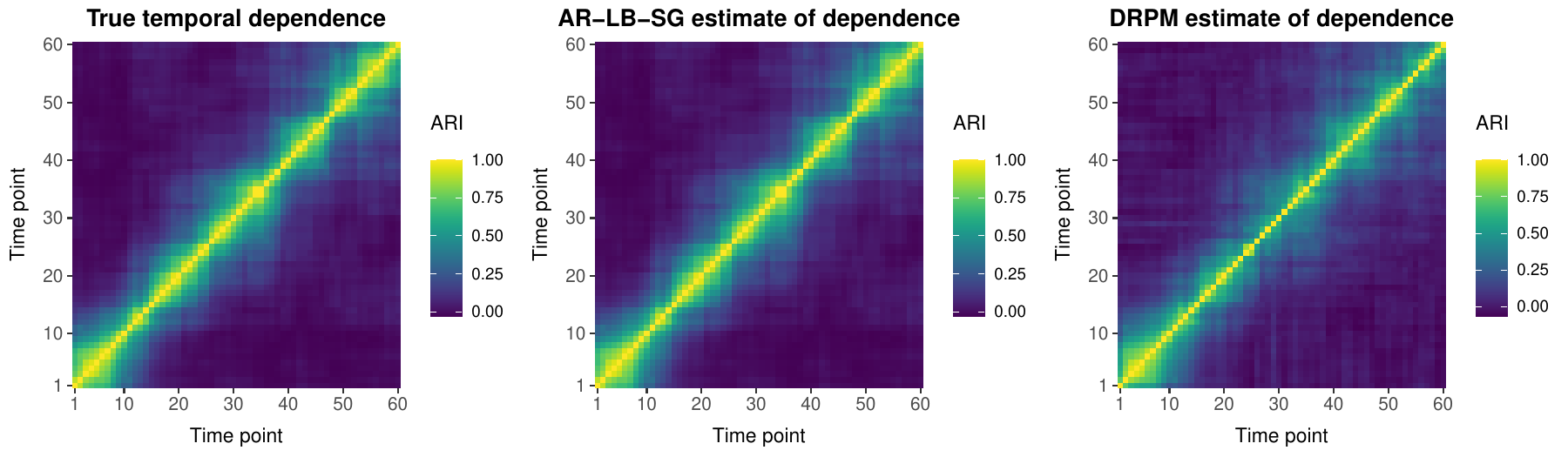}

\vspace{0.2cm}

\includegraphics[width=0.9\linewidth]{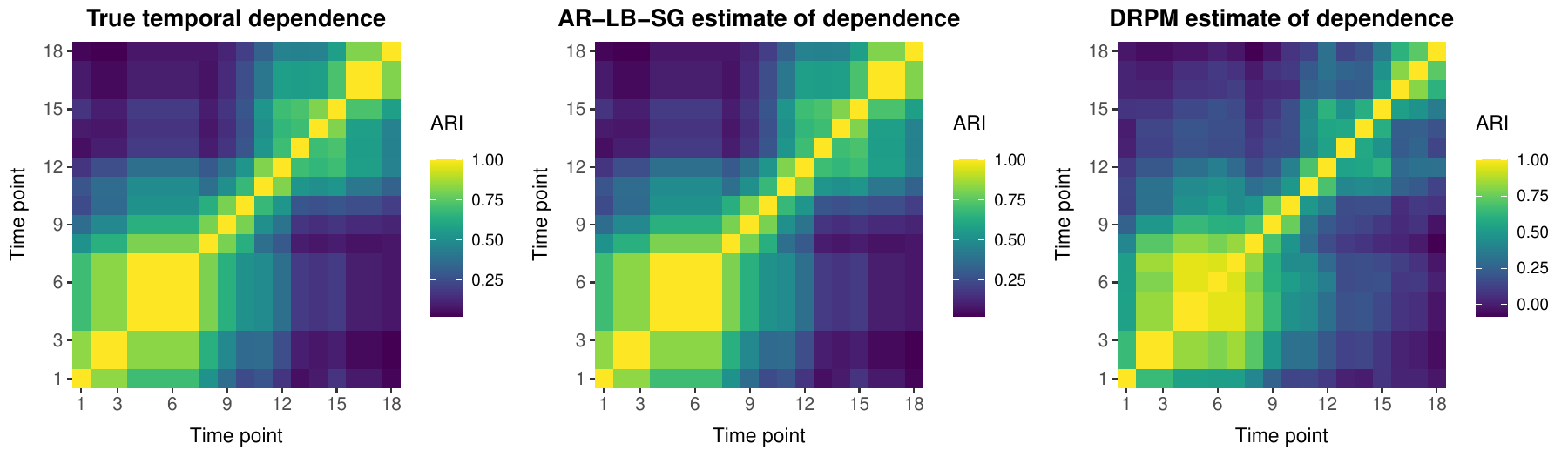}
\end{center}
\caption{True temporal dependence, measured by lagged ARI values, and the estimated temporal dependence recovered by the AR-LB-SG process and the DRPM. The first row presents the temporal dependence across the entire 60 time points, while the second row presents the temporal dependence for the first 18 time points. At each time point, partitions were estimated by minimizing the VI loss function.}
\label{fig:temporal_recovery_sim1}
\end{figure*}

Simulation results from a single preliminary run---under a partially unbalanced clusters case---are presented in Figures \ref{fig:coclust_sim1}, \ref{fig:hists_sim1}, and \ref{fig:temporal_recovery_sim1}. Particularly, Figure \ref{fig:coclust_sim1} presents the true co-clustering structures and their corresponding posterior co-clustering probabilities recovered by the AR-LB-SG process and the DRPM at three different time points. Figure \ref{fig:hists_sim1} presents histograms of the simulated data and their corresponding posterior predictive distributions recovered by the two methods. Lastly, Figure \ref{fig:temporal_recovery_sim1} presents the true temporal dependence between the partitions, as well as the estimated temporal dependence recovered by the AR-LB-SG process and the DRPM. As in \cite{page2022dependent}, we measure temporal dependence using a time-lagged adjusted rand index (ARI, \cite{hubert1985comparing}). To do so, we first obtain point estimates of the partitions at each time point and then compute ARI values for each pair of partitions separated by a given lag. Point estimates of the partitions are obtained by minimizing the posterior expectation of variation of information (VI) loss function \citep{wade2018bayesian} through the \textbf{R} package "\texttt{BNPmix}" \citep{corradin2021bnpmix}. We employ the VI loss function because it is a proper metric on the partition space and it provides a more accurate recovery of latent partitions compared to other loss functions, such as \textit{Binder’s} loss function \citep{BINDER_Loss}---which tends to overestimate the number of clusters \citep{wade2018bayesian}. ARI values are then computed using the \textbf{R} package "\texttt{salso}" \citep{salso}.

Note, from Figure \ref{fig:coclust_sim1}, that even though the DRPM produces more decisive posterior co-clustering probabilities---leaning strongly toward zero or one rather than remaining near 0.5---our proposed method consistently achieves a co-clustering structure that is closer to the true one, while still maintaining decisive probabilities. Interestingly, though, Figure \ref{fig:hists_sim1} shows that both methods tend to produce similar posterior predictive distributions. Lastly, from Figure \ref{fig:temporal_recovery_sim1}, we have that the AR-LB-SG process captures better the true temporal dependence between the partitions. Similar plots obtained under a balanced clusters case are presented in the Supplementary Materials, which display consistent results.
\par
That being said, the results presented in Figures \ref{fig:coclust_sim1}, \ref{fig:hists_sim1}, and \ref{fig:temporal_recovery_sim1} are based on a single preliminary run. To better understand the differences between the AR-LB-SG process and the DRPM, this experiment is repeated $S=200$ different times under balanced and partially unbalanced clusters. As comparison metrics, we employ the Watanabe--Akaike information criterion (WAIC), the Pareto-smoothed importance sampling leave-one-out cross-validation (PSIS-LOO, \cite{vehtari2017practical}), Binder's loss function between the true partitions and the estimated ones, the VI loss function, the mean squared error of the estimated $\theta_{it}$'s, the distance between the true co-clustering matrices and the matrices of posterior co-clustering probabilities, and the elapsed (wall-clock) time in minutes. WAIC and PSIS-LOO are implemented through the \textbf{R} package "\texttt{loo}" \citep{loo_pckg}. To compute Binder's loss and the VI loss functions, we make use of the \textbf{R} package "\texttt{salso}" \citep{salso}. We employ the posterior mean as a point estimate of $\theta_{it}$, and, for $t\in\{1,\dots,T\}$, let $\lVert \mathbf{C}_t - \hat{\mathbf{C}}_t \rVert_{F}$ be the Frobenius norm of the difference between the true co-clustering matrix, $\mathbf{C}_t$, and the recovered matrix of posterior co-clustering probabilities, $\hat{\mathbf{C}}_t$. Then, we define the average co-clustering error as $T^{-1}\sum_{t=1}^{T}\lVert \mathbf{C}_t - \hat{\mathbf{C}}_t \rVert_{F}$. Additional details on the implementation of WAIC and PSIS-LOO are presented in the Supplementary Materials.

\begin{figure*}
\begin{center}
\includegraphics[width=\linewidth]{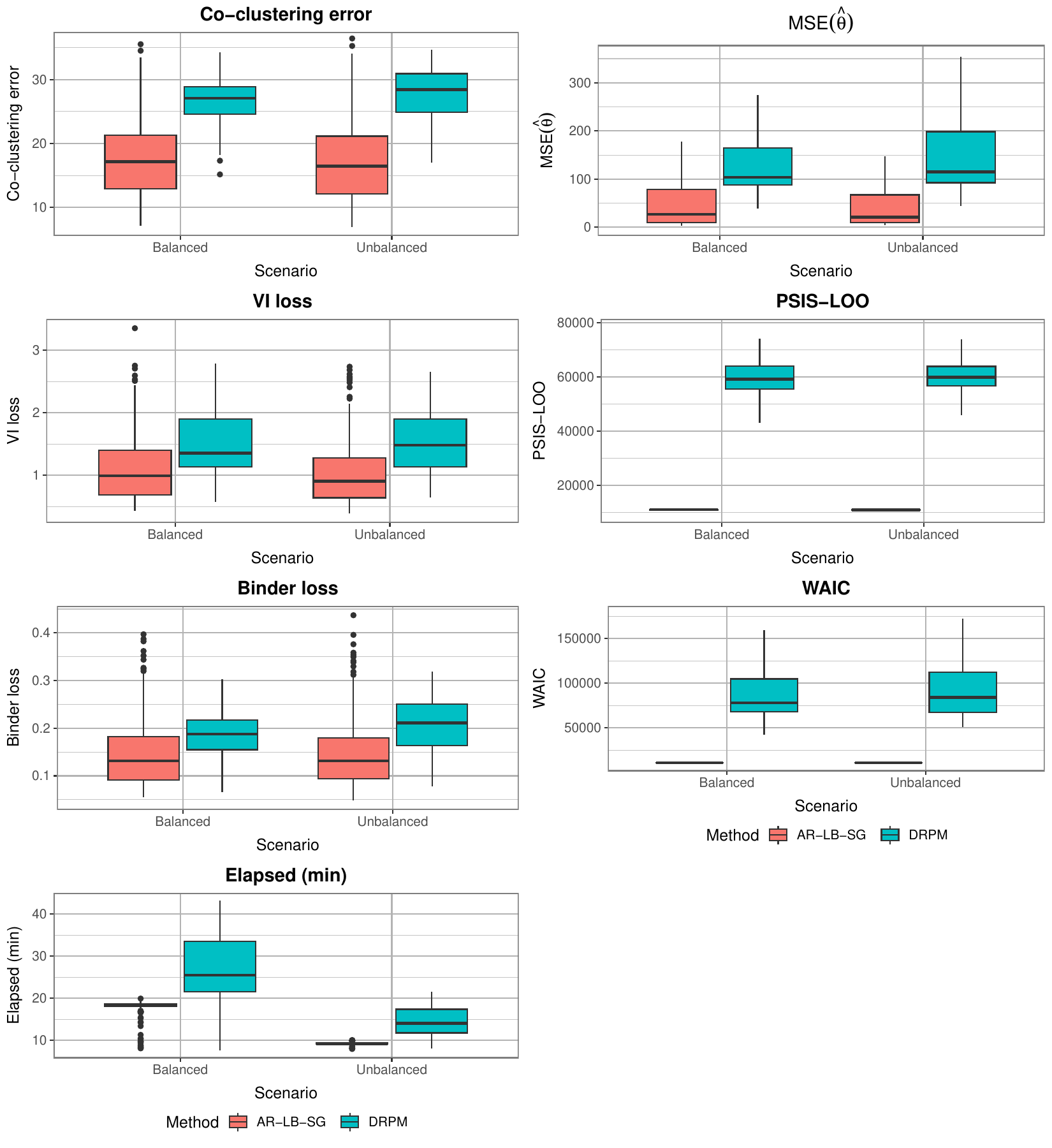}
\end{center}
\caption{Distributions of the comparison metrics obtained with the AR-LB-SG process and the DRPM based on 200 independent replications. In all cases, lower values indicate a better performance.}
\label{fig:boxplots_sim1}
\end{figure*}
\par
Figure \ref{fig:boxplots_sim1} presents the distributions of the comparison metrics obtained with the AR-LB-SG process and and the DRPM, based on 200 independent replications. In all cases, lower values indicate a better performance. It is clear, then, that the AR-LB-SG process is at least comparable and often outperforms the DRPM across all metrics. In particular, the smaller WAIC and PSIS-LOO values indicate that the AR-LB-SG process provides a better model fitting without incurring in notable overfitting issues. The smaller Binder's, VI, and co-clustering errors from the AR-LB-SG process highlight its capability of performing dynamic clustering.  Lastly, the smaller elapsed times demonstrate that our proposed method can be easily deployed in practice and can be executed on a personal laptop.
\par
On the whole, these results suggest that the AR-LB-SG process is a strong competitor in the field of Bayesian nonparametric dynamic clustering. To make our method more accessible to a wider range of practitioners, we implement it in the \textbf{R} package "\texttt{dynclusts}", included in the Supplementary Materials. 

\section{Application to Chilean FSP data}
\label{sec:chile_data}

We now make use of the AR-LB-SG process in order to identify dynamic clusters of air-quality monitoring stations exhibiting similar FSP levels across continental Chile, which could then guide targeted interventions, localized public-health response measures, and regional mitigation strategies.
\par
In particular, we employ data recorded across 64 monitoring stations (depicted in Figure \ref{fig:stations}) from January 2020 to December 2024 (i.e., 60 months). More formally, let $Y_{it}$ be the monthly average concentration level of FSP (in $\mu \text{g m}^{-3}$) recorded in monitoring station $i$ at time $t$. As covariates, we consider, for station $i$ at time $t$, the relative humidity (as a percentage of the maximum possible water vapor in the air at a given temperature), the wind direction (in angular degrees), the wind speed (in meters per second), and the square of humidity. To account for the circular nature of the wind direction, we split the recorded variable into its sine and cosine components. More formally, let $\texttt{windDir}$ be the recorded wind direction. We then split $\texttt{windDir}$ into two variables, namely, (1) $\texttt{sin\_windDir} = \sin\left(\texttt{windDir} \times \frac{\pi}{180}\right)$ and (2) $\texttt{cos\_windDir} = \cos\left(\texttt{windDir} \times \frac{\pi}{180}\right)$, where $\pi$ denotes the well-known mathematical constant (i.e., $\pi\approx 3.14159$). To facilitate reproducibility, additional details about the data preparation process---as well as the data themselves and all the source code used to prepare the data---are all included in the Supplementary Materials. 
\par
\begin{figure}
\begin{center}
    \includegraphics[width=0.8\linewidth]{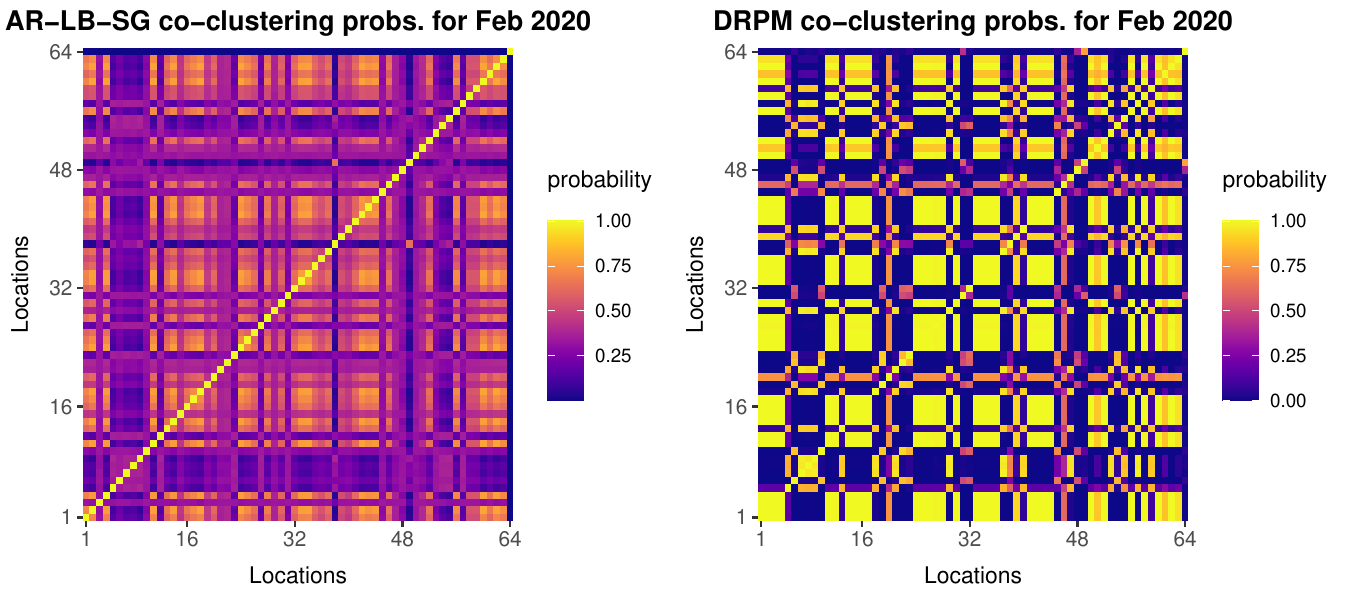}
    \includegraphics[width=0.8\linewidth]{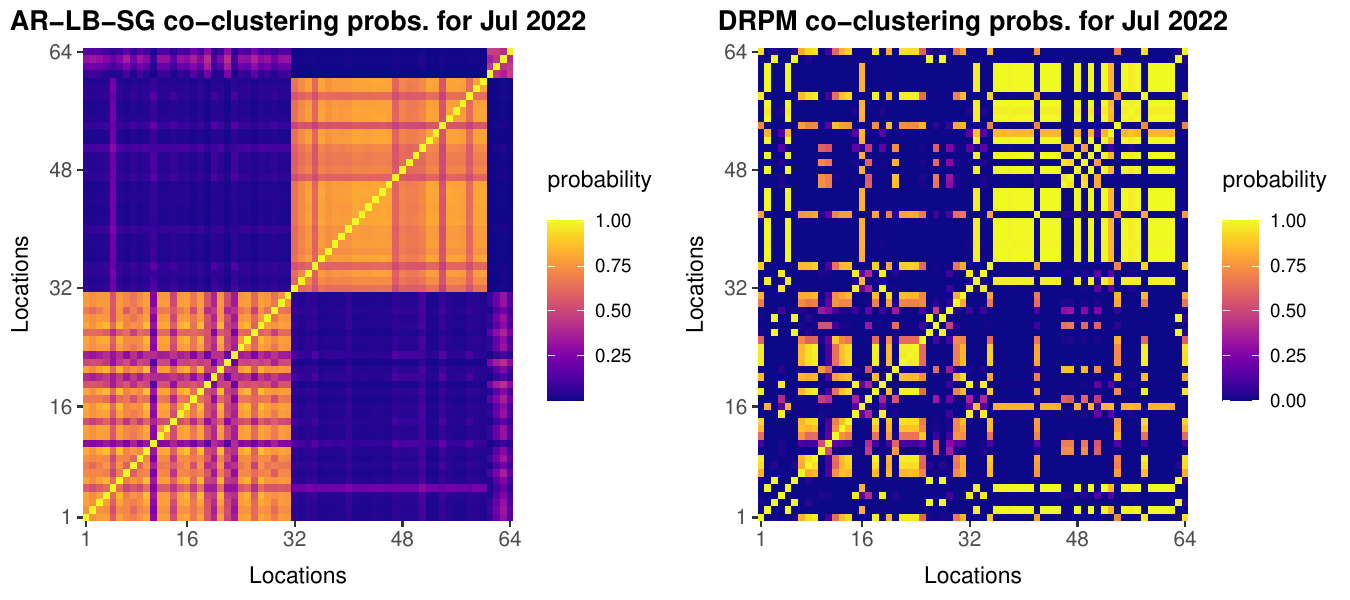}
    \includegraphics[width=0.8\linewidth]{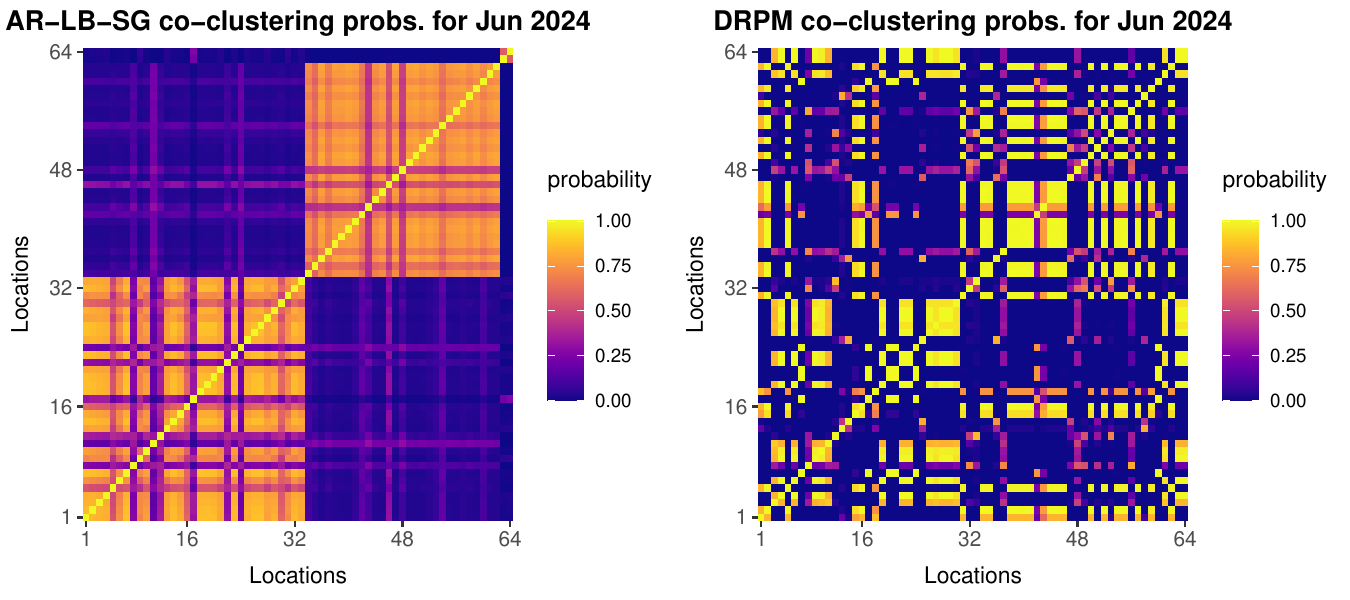}
    \includegraphics[width=0.8\linewidth]{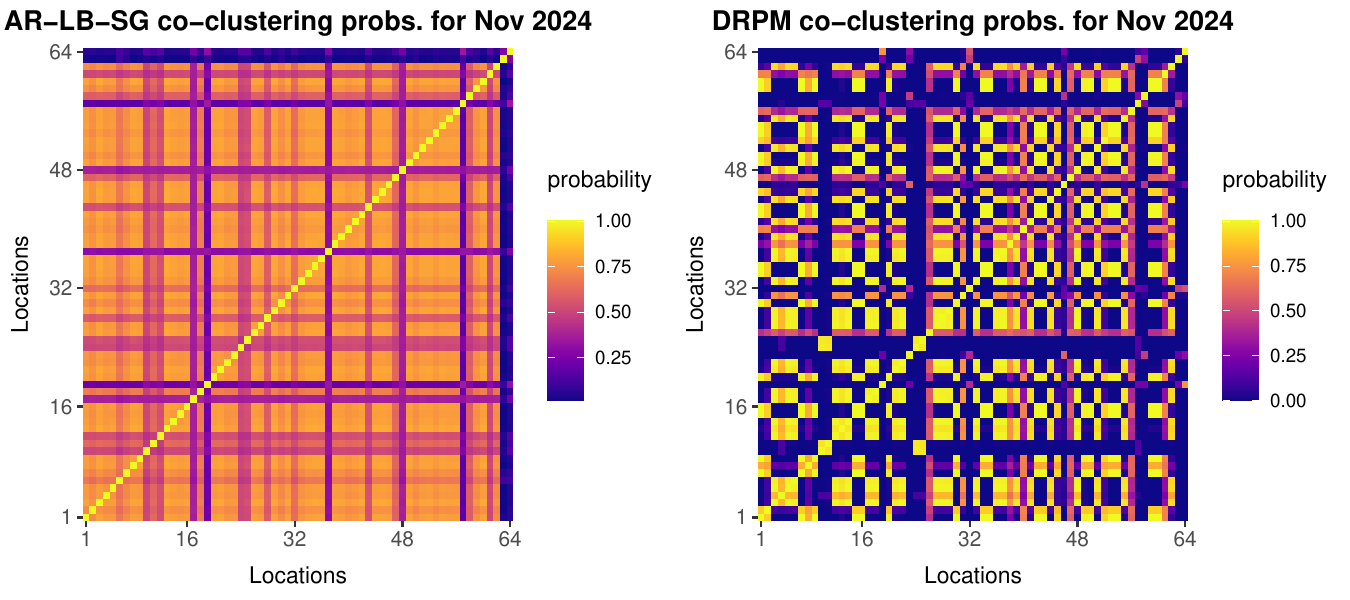}
\end{center}
\caption{Posterior co-clustering probabilities recovered by the AR-LB-SG process and the DRPM at four different time points. Results are based on the Chilean FSP data.}
\label{fig:chile_coclust_months}
\end{figure}
\begin{figure}
\begin{center}
    \includegraphics[width=0.39\linewidth]{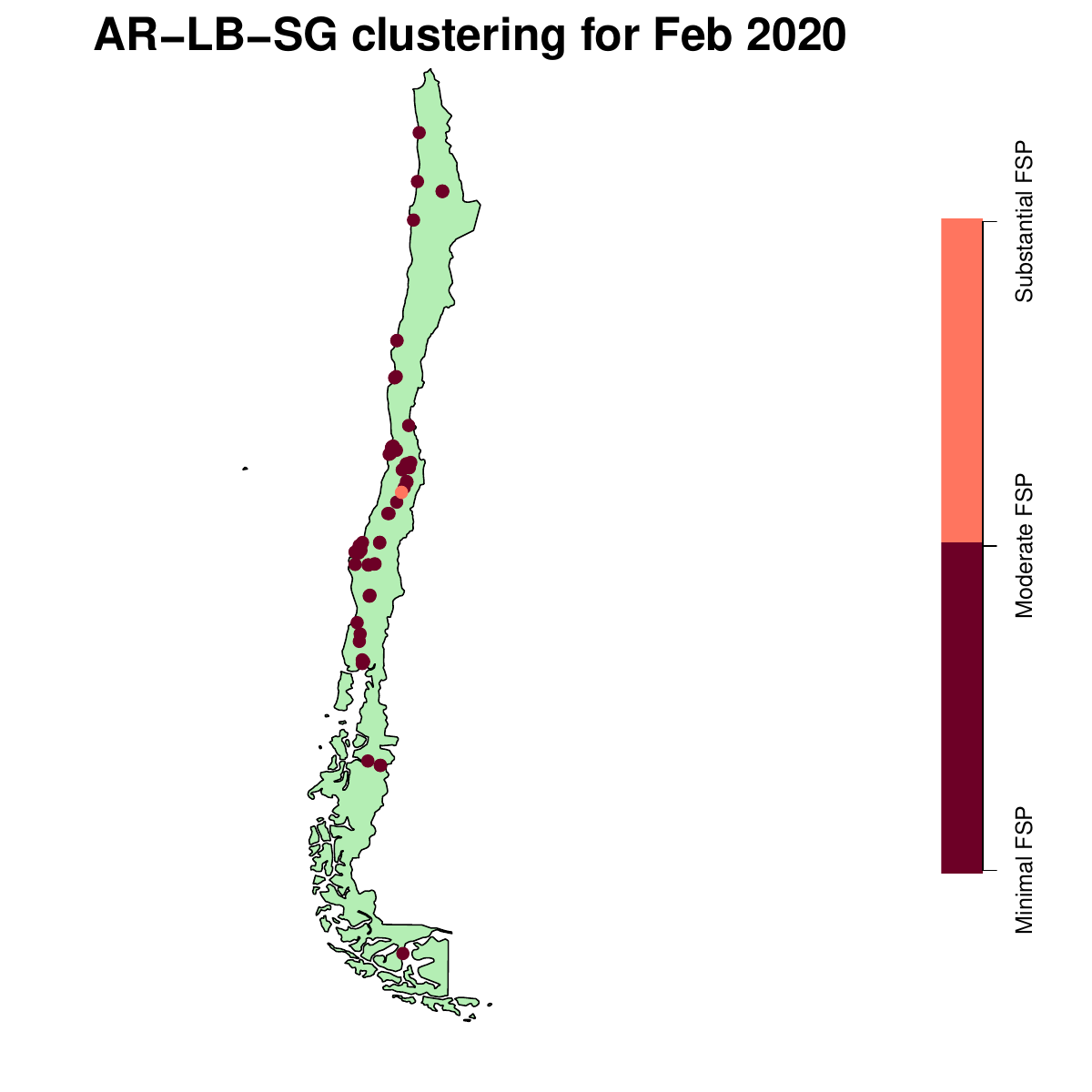}
    \includegraphics[width=0.39\linewidth]{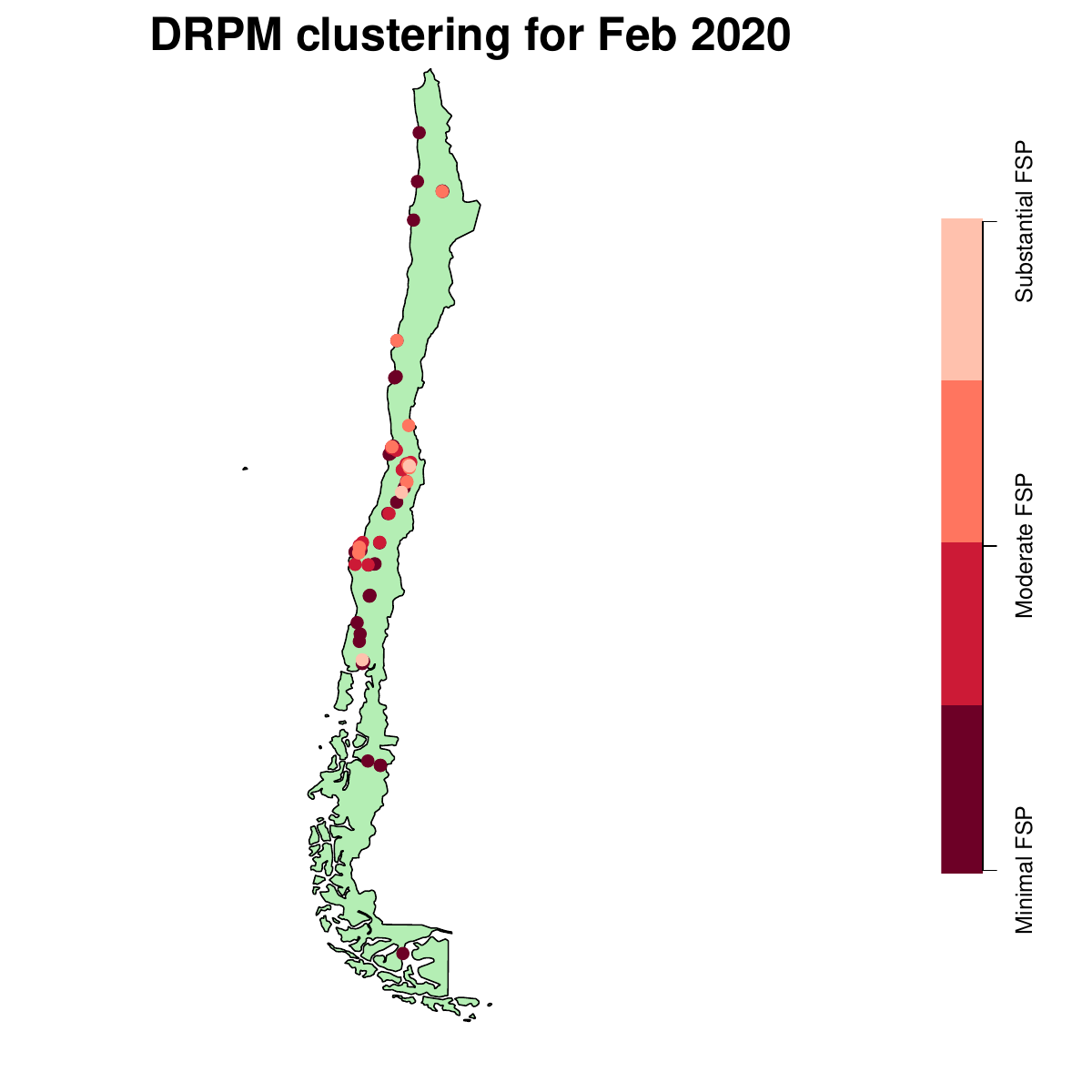}
    \includegraphics[width=0.39\linewidth]{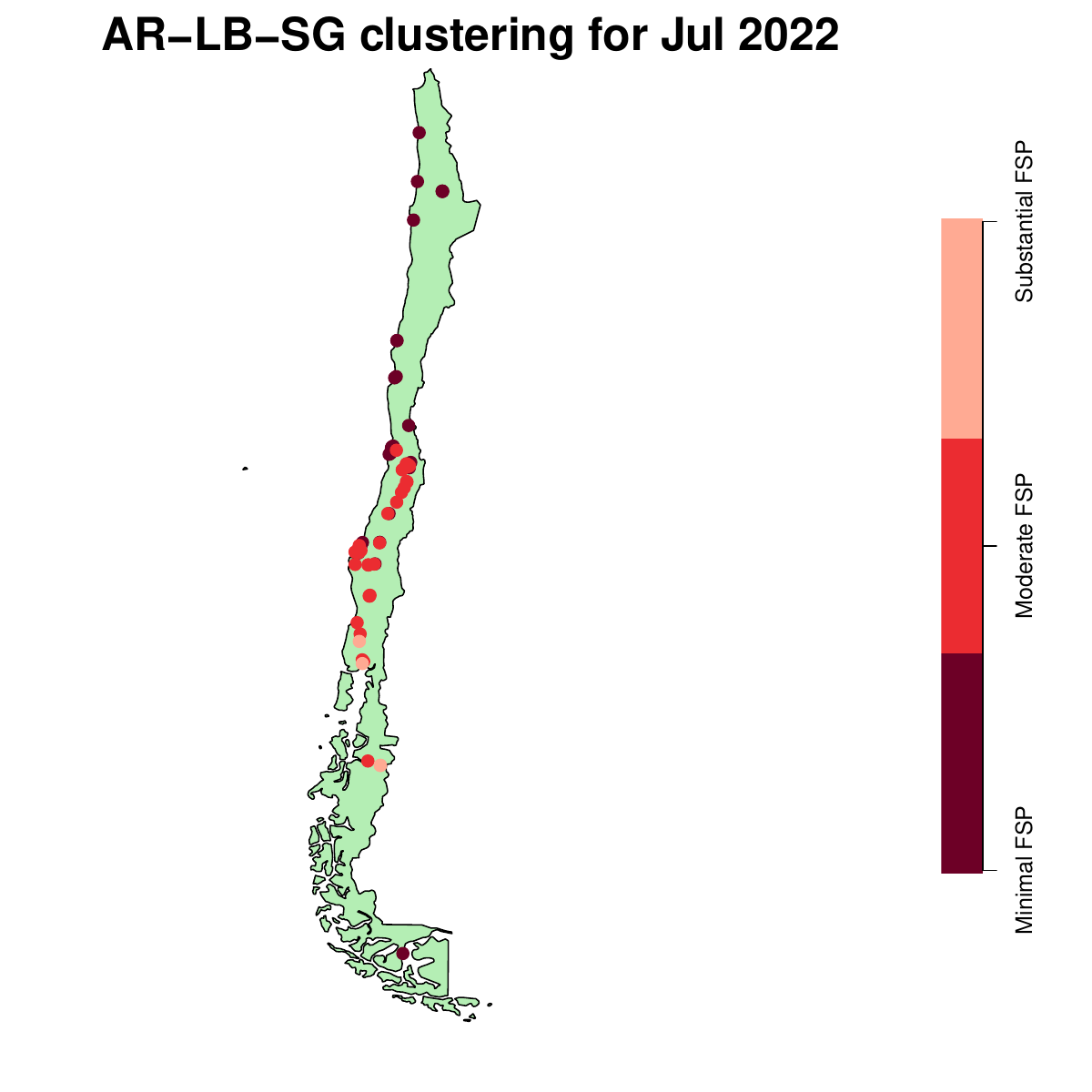}
    \includegraphics[width=0.39\linewidth]{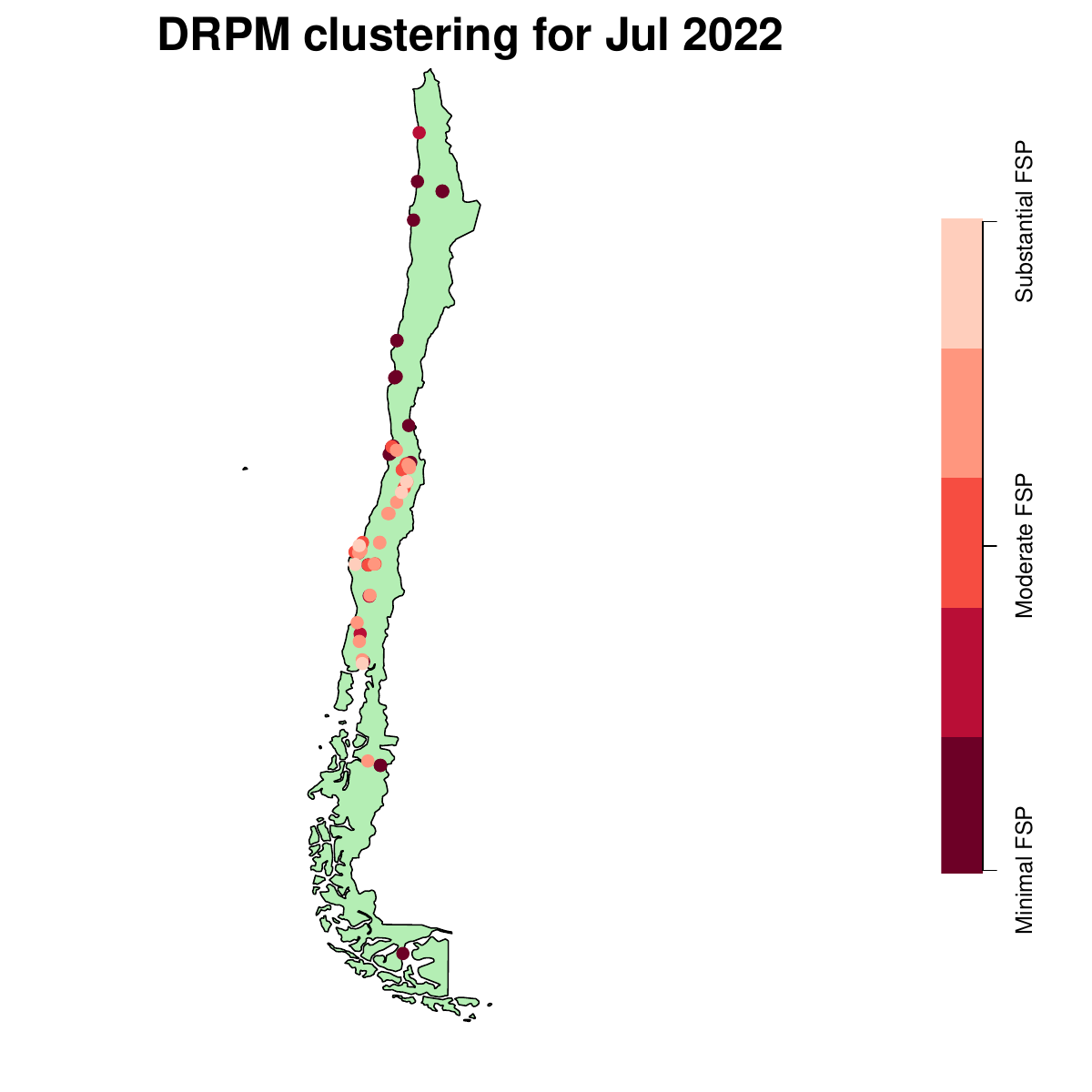}
    \includegraphics[width=0.39\linewidth]{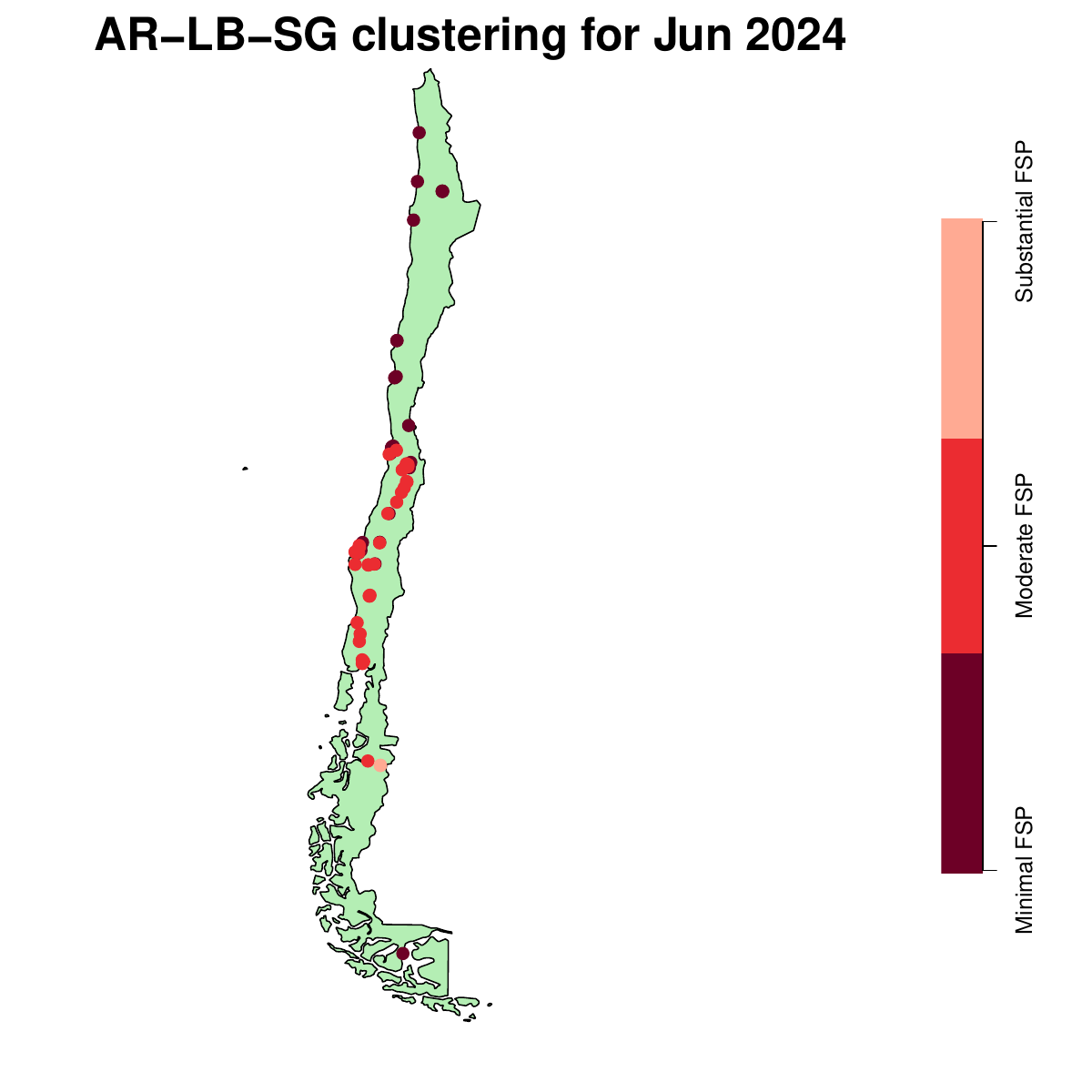}
    \includegraphics[width=0.39\linewidth]{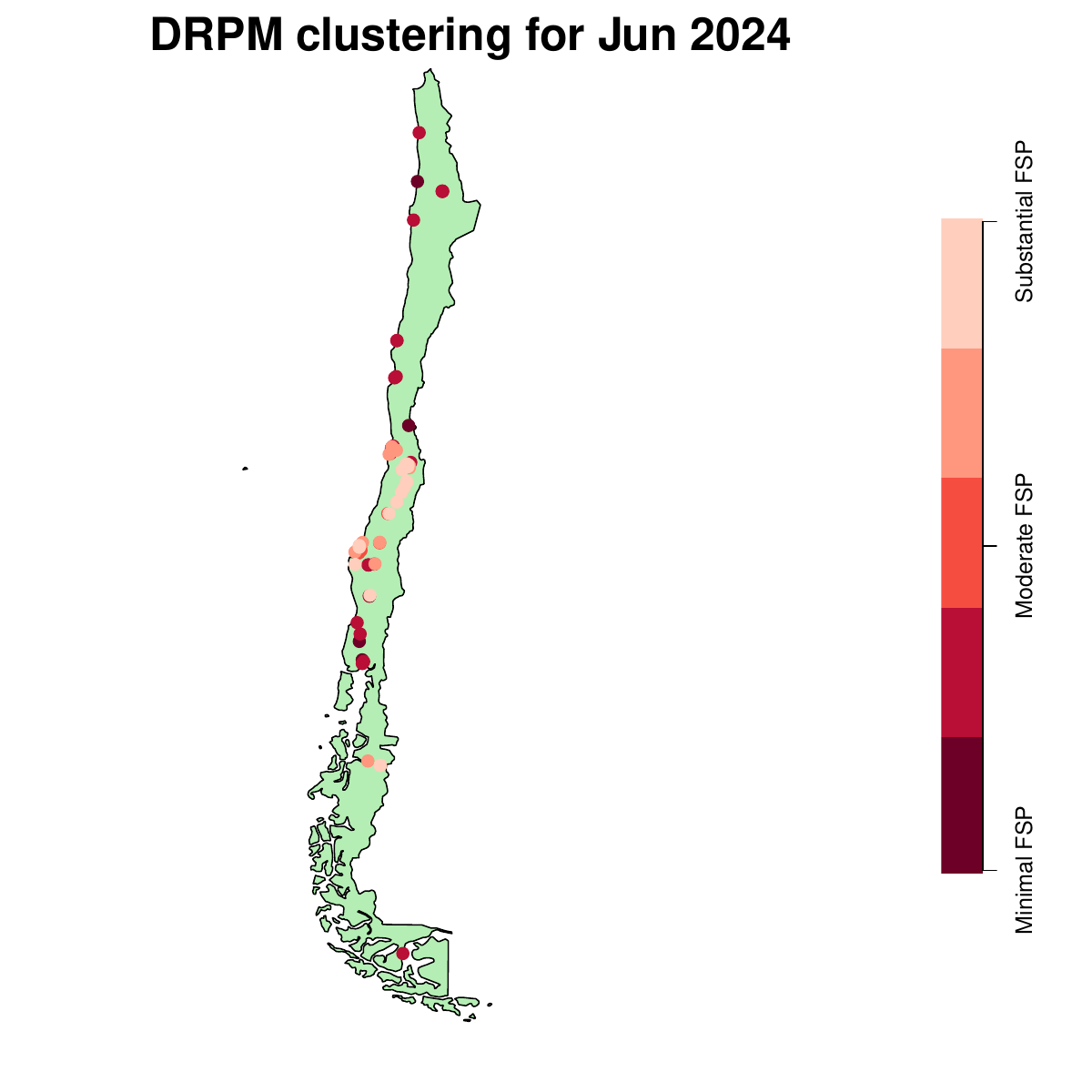}
    \includegraphics[width=0.39\linewidth]{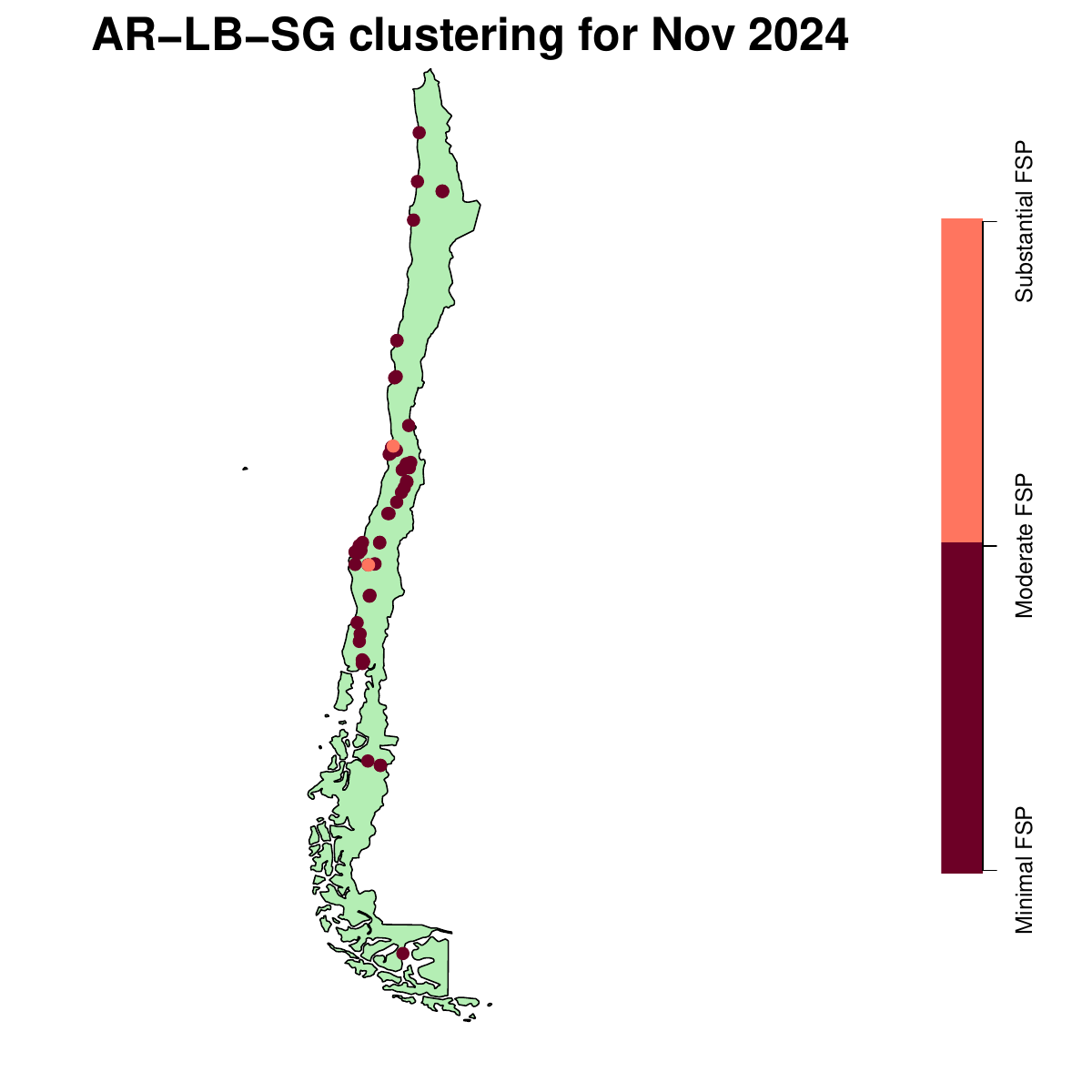}
    \includegraphics[width=0.39\linewidth]{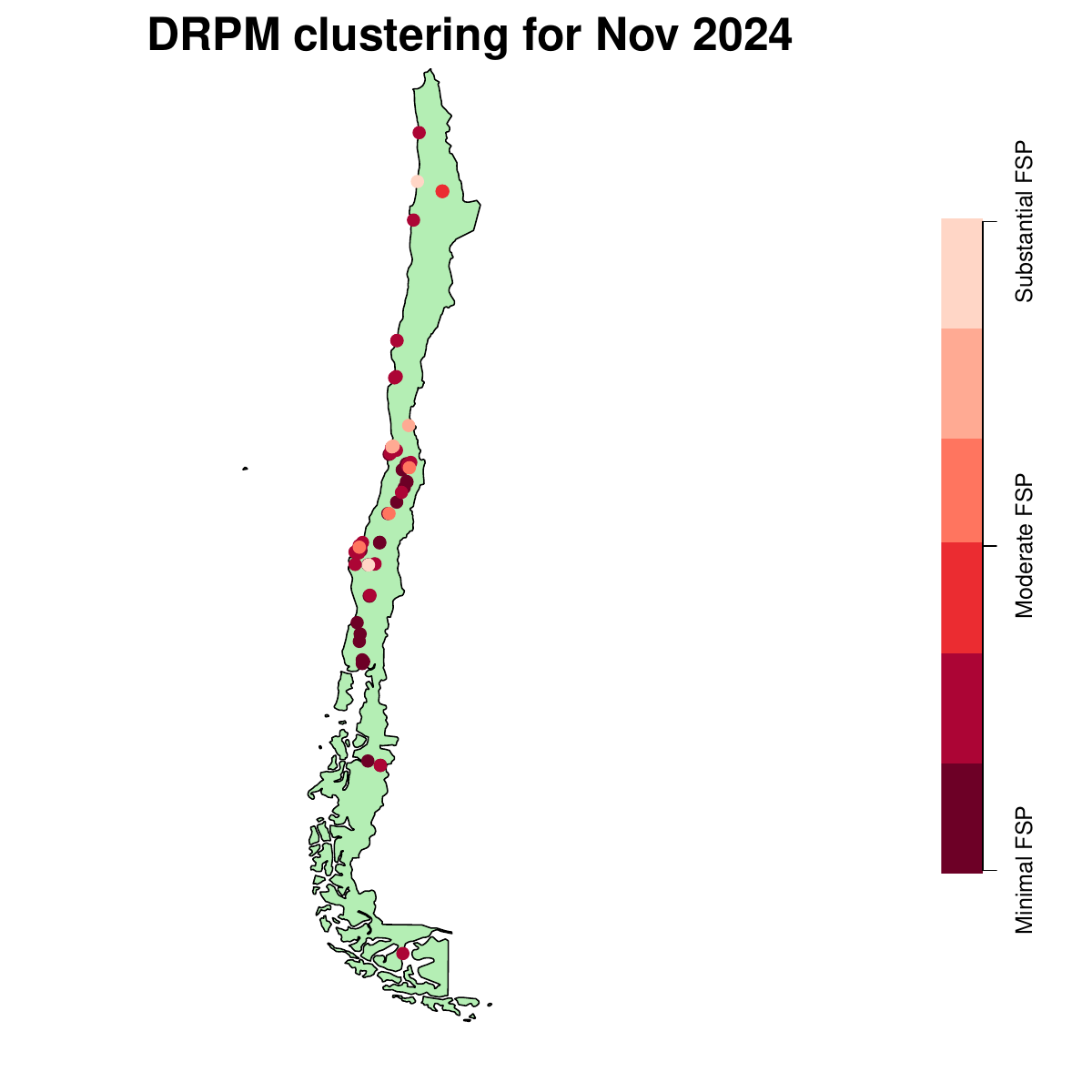}
\end{center}
\caption{Partitions—and their corresponding geographical locations—recovered by the AR-LB-SG process and the DRPM at four different time points. Results are based on the Chilean FSP data.}
\label{fig:chile_maps_results}
\end{figure}
\begin{figure*}
\begin{center}
\includegraphics[width=0.8\linewidth]{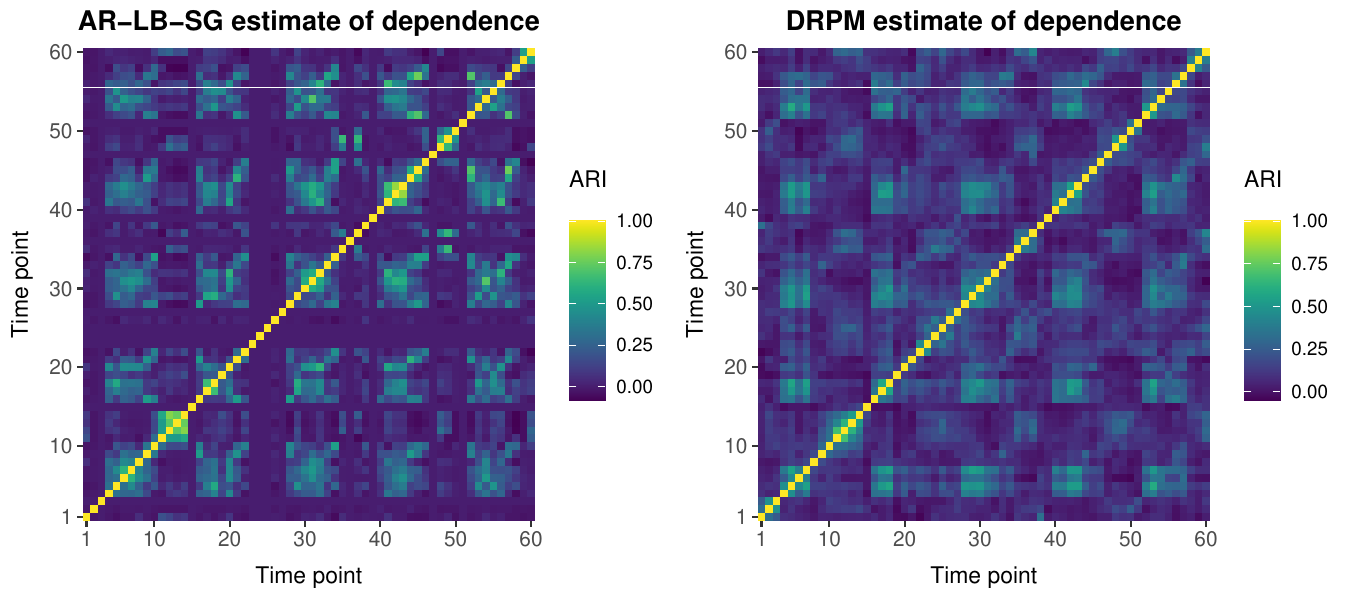}
\end{center}
\caption{Estimated temporal dependence, measured by lagged ARI values, recovered by the AR-LB-SG process and the DRPM. At each time point, the partition was estimated by minimizing the VI loss function. Results are based on the Chilean FSP data.}
\label{fig:chile_temporal_ari}
\end{figure*}

As in Section \ref{sec:simulations}, we fit the AR-LB-SG process and the DRPM to these data. In both cases, we run the MCMC algorithms for 200000 iterations, discard the first 20000 draws as burn-in, and thin every $25^{\textnormal{th}}$ draw---for a total of 7200 retained MCMC samples. All remaining hyperparameter settings are the same as those described in Section \ref{sec:simulations}. Results from the analysis are presented in Table \ref{tab:chileFSP}, as well as in Figures \ref{fig:chile_coclust_months}, \ref{fig:chile_maps_results}, and \ref{fig:chile_temporal_ari}. 

\begin{table*}
\caption{Model assessment based on the Chilean FSP data. \\ The best model, across each metric, is highlighted in bold.}
\label{tab:chileFSP}
\begin{tabular}{@{}lccc}
\hline
Model & \multicolumn{1}{c}{WAIC}
& \multicolumn{1}{c}{PSIS-LOO} & \multicolumn{1}{c}{Elapsed (min)} \\
\hline
AR-LB-SG & \textbf{23972.75} & \textbf{25171.43} & \textbf{67.75} \\
DRPM     & 781705.60          & 157016.40          & 119.80 \\
\hline
\end{tabular}
\end{table*}

Table \ref{tab:chileFSP} reports the WAIC, PSIS-LOO, and the elapsed times from the two competing models, when applied to the Chilean FSP data. The notably smaller WAIC and PSIS-LOO values from the AR-LB-SG process illustrate its superior fit the Chilean FSP data over the DRPM. We can also observe that the AR-LB-SG process is around 44\% faster than the DRPM, demonstrating its strong computational efficiency.

Figure \ref{fig:chile_coclust_months} presents the posterior co-clustering probabilities recovered by the AR-LB-SG process and the DRPM at four different time points. Interestingly, we can observe that during warmer months in the Southern Hemisphere, such as November and February, the AR-LB-SG process tends to group most monitoring stations into a single cluster. In contrast, during colder months, particularly June and July, the AR-LB-SG process tends to identify two main clusters of monitoring stations. These results are consistent with the histograms presented in Figure \ref{fig:chile_hists}, which illustrate stronger homogeneity during warmer months and heterogeneity during colder ones. Lastly, it is clear that the DRPM tends to produce less parsimonious and interpretable results, which is also evident in its larger WAIC and PSIS-LOO values.
\par
To better visualize this phenomenon, Figure \ref{fig:chile_maps_results} presents point estimates of the partitions of monitoring stations identified by the AR-LB-SG process and the DRPM. As in Section \ref{sec:simulations}, we obtain point estimates of the partitions by minimizing the VI loss function. It is clear, then, that during warmer months, the AR-LB-SG process tends to group most monitoring stations within a single cluster, while signaling the existence of a few singletons. For instance, in February 2020, we can observe a singleton station in the central part of the country. Such a singleton station is ``San Fernando,'' in the region of ``El Liberatdor General Bernarndo O'Higgins,'' which recorded a monthly average concentration of FSP of 127.5 $\mu \text{g m}^{-3}$. Interestingly, though, such a reading is labeled as \textit{preliminary} rather than \textit{validated}. This suggests that the AR-LB-SG process can identify anomalous stations, which can then guide better management and quality-assurance strategies.
\par 
It is also clear that, during colder months, the AR-LB-SG process identifies clusters of monitoring stations exhibiting low FSP concentration levels in the northern part of the country and clusters of monitoring stations exhibiting substantial FSP concentration levels in the south. These results are not surprising either, as the northern part of the country is closer to the Earth's tropics and therefore experiences milder winters. In contrast, the southern regions experience more severe winters, leading to a greater demand for winter heating.  It is well-known that winter heating is one of the major drivers of FSP concentration levels (see e.g., \cite{liang2015assessing}). As such, our results are in line with the existing literature. The DRPM, on the other hand, is consistently recovering a larger number of clusters, resulting in less interpretable results. Combined with its notably larger WAIC and PSIS-LOO values, this suggests that the DRPM is more prone to overfitting issues. 

Lastly, Figure \ref{fig:chile_temporal_ari} presents the temporal dependence, measured by lagged ARI values, recovered by the AR-LB-SG process and the DRPM. It is clear that both methods are identifying partitions with a seasonal pattern, which is consistent with our previous results (see e.g., Figures \ref{fig:chile_coclust_months} and \ref{fig:chile_maps_results}). Note, however, that such a seasonal pattern is much more evident under the AR-LB-SG process. On the whole, this reinforces the idea that seasonal events, like winter heating, may drive latent clusters of air-quality monitoring stations exhibiting similar FSP levels across continental Chile, even after controlling for various covariates. That being said, additional research is still needed to confirm and clarify the nature of these dependency links. This, however, is outside the scope of this article.

\section{Discussion}
\label{sec:discussion}
Fine suspended particulates, commonly known as $\textnormal{PM}_{2.5}$, are among the most harmful air pollutants, posing serious risks to population health and environmental integrity. Despite these well-known adverse effects, many regions worldwide continue to experience FSP concentration levels that substantially exceed recommended health standards. One notable example is the southern part of the Andean region, where FSP levels frequently remain well above safe limits. Therefore, throughout this article, we have addressed the challenge of accurately identifying clusters of air-quality monitoring stations exhibiting similar FSP concentration levels across continental Chile. To account for the notable dependencies of FSP concentration levels on various regional and temporal factors, we have capitalized on the flexibility of Bayesian nonparametric methods. In particular, we have built upon the ideas of dynamic clustering from \cite{DeIorio2023_AR_DPMs}, in which clustering structures may be influenced by complex temporal and regional dependencies. That being said, existing implementation of dynamic clustering rely on copula-based dependent DPs, which, despite their versatility, remain computationally impractical for real-world deployment.
\par
With this in mind, we have generalized the novel idea of logistic-beta dependent DPs \citep{logistic_beta_2025}, a more efficient alternative to copula-based dependent DPs, in order to perform dynamic clustering. Furthermore, to facilitate the process of eliciting an informative prior distribution and make the workflow more transparent and straightforward, we have also incorporated a Stirling-gamma prior, a novel distribution family, on the concentration parameter of our underlying DP \citep{zito2023bayesian}. As such, we have called our proposed method the autoregressive logistic-beta Stirling-gamma process.
Lastly, to make our contributions widely available to the scientific community, we have implemented our proposed method in the open-source \textbf{R} package "\texttt{dynclusts}", included in the Supplementary Materials. 
\par
When applied to the Chilean FSP data, our proposed method reveals an intriguing seasonal pattern in the clustering structure of air-quality monitoring stations. More precisely, during warmer months, our method tends to cluster together the majority of the monitoring stations---with very few exceptions that could be considered anomalous stations---suggesting that after controlling for various covariates and incorporating location-specific random effects, there are no major differences between FSP levels across stations. During colder months, on the other hand, we can observe clusters of monitoring stations exhibiting low FSP concentration levels in the northern part of the country and clusters of monitoring stations exhibiting substantial FSP concentration levels in the south, which suggests that seasonal events, such as winter heating \citep{liang2015assessing}, may have a strong influence on FSP clusters across continental Chile, even after accounting for other variables. These results can then aid public health and environmental authorities determine where and when considerable pollution episodes occur, which could then guide targeted interventions, localized public-health responses, environmental surveillance, and better management strategies. Lastly, we have also illustrated the superior performance of our proposed method over state-of-the-art algorithms for dynamic clustering, like the dependent random partition model from \cite{page2022dependent}. 
\par
The results from this article, however, also raise extensions and future challenges beyond the scope of this paper. For instance, one could study in greater detail the dependency links between seasonal events---like winter heating---and the recovered partitions. Additionally, as discussed in \cite{logistic_beta_2025}, the logistic-beta process can also be used to induce dependencies in other discrete random probability measures like the Pitman--Yor process \citep{perman1992size, pitman1997PDP} or even more general stick-breaking processes as in \cite{Ishwaran_James_2001_stick}.  As such, another natural extension could be to consider a broader class of discrete random measures, beyond the DP, in order to perform dynamic clustering.  One last extension could be to consider a \textit{hierarchical dynamic clustering}, in the spirit of \cite{Teh01122006}, in which practitioners observe data from multiple countries---such as Chile, Brazil, and Argentina---aiming to identify dynamic pollution clusters within each country, while allowing some degree of information sharing across countries. Therefore, our proposed method has also introduced promising avenues for subsequent research. More broadly, though, this article has expanded and enriched the already vibrant world of dependent DPs, as well as the emerging field of Bayesian nonparametric dynamic clustering.



\begin{supplement}
\stitle{Supplementary Materials}
\sdescription{Derivations, proofs, additional results, and supplemental information referenced in the article (\texttt{.pdf} file).}
\end{supplement}
\begin{supplement}
\stitle{dynclusts}
\sdescription{The "\texttt{dynclusts}" \textbf{R} package, which implements Bayesian nonparametric dynamic clustering through an autoregressive logistic-beta Stirling-gamma process as described in this article (\texttt{.zip} file). The "\texttt{dynclusts}" \textbf{R} package is also available online at \url{https://github.com/marinsantiago/dynclusts}}
\end{supplement}
\begin{supplement}
\stitle{dynclusts-applications}
\sdescription{\textbf{R} code and data to reproduce the results from this article (\texttt{.zip} file). Source code and data are available at \url{https://github.com/marinsantiago/dynclusts-applications}}  
\end{supplement}


\bibliographystyle{imsart-nameyear} 
\bibliography{citation}       

\end{document}